\documentclass[format=acmsmall, review=false, screen=true]{acmart}

\usepackage{booktabs} % For formal tables
\usepackage{tabularx}
\usepackage{enumitem}
\usepackage{multirow}
\usepackage{mdframed}
\usepackage{soul,color}

\usepackage[ruled]{algorithm2e} % For algorithms

\SetAlFnt{\small}
\SetAlCapFnt{\small}
\SetAlCapNameFnt{\small}
\SetAlCapHSkip{0pt}
\IncMargin{-\parindent}

\acmJournal{CSUR}
\acmVolume{0}
\acmNumber{0}
\acmArticle{0}
\acmYear{0000}
\acmMonth{0}
\copyrightyear{0000}
\setcopyright{acmlicensed}
\acmDOI{0000001.0000001}

\received{July 2021}
\newcommand{\rev}[1]{{\color{black}#1}}

\begin{document}
% Title portion. Note the short title for running heads
\title[Survey on Formal Methods for Railways]{Formal Methods in Railways: a Systematic Mapping Study}

\author{Alessio Ferrari}
\orcid{0000-0002-0636-5663}
\email{alessio.ferrari@isti.cnr.it}
\author{Maurice H. ter Beek}
\orcid{0000-0002-2930-6367}
\email{maurice.terbeek@isti.cnr.it}
\affiliation{%
  \institution{Istituto di Scienza e Tecnologie dell'Informazione, Consiglio Nazionale delle Ricerche (ISTI--CNR)}
  \streetaddress{Via G. Moruzzi, 1}
  \city{Pisa}
  \postcode{56124}
  \country{Italy}
}

\begin{abstract}
Formal methods are mathematically based techniques for the rigorous development of software-intensive systems. The railway signaling domain is a field in which formal methods have traditionally been applied, with several success stories. 
%The railway field is particularly important today with the transition towards green transportation, and formal methods are strategically crucial to guarantee dependability in the face of increasing complexity of these systems. 
This article reports on a mapping study that surveys the landscape of research on applications of formal methods to the development of railway systems. 
Following the guidelines of systematic reviews, we identify 328 relevant primary studies, and extract information about their demographics, the characteristics of formal methods used and railway-specific aspects. 
%and analyze them according to three main research questions, concerning the state of the literature from a demographic and empirical viewpoint, the characteristics of formal methods applied, and railway-specific aspects. Furthermore, we also consider two cross-cutting questions, to identify industrial studies, and  trends of recent years.  
Our main results are as follows: (i)~we identify a total of 328 primary studies relevant to our scope published between 1989 and 2020, of which 44\% published during the last 5 years and 24\% involving industry; (ii)~the majority of  studies are evaluated through Examples~(41\%) and Experience Reports~(38\%), while full-fledged Case Studies are limited~(1.5\%); (iii)~Model checking is the most commonly adopted technique~(47\%), followed by simulation~(27\%) and theorem proving~(19.5\%); 
%Less commonly used are techniques that are strictly related to code, as test or code generation; 
(iv)~the dominant languages are UML~(18\%) and B~(15\%), while frequently used tools are ProB~(9\%), NuSMV~(8\%) and UPPAAL~(7\%); however, a diverse landscape of languages and tools is employed; (v)~the majority of systems are interlocking products~(40\%), followed by models of high-level control logic~(27\%);
%, in some cases with reference to ERTMS-ETCS (19\%), CTCS (5\%), and CBTC (8\%) standards; 
(vi)~most of the studies focus on the Architecture~(66\%) and Detailed Design~(45\%) development phases. 
%Our results on recent trends show that the field is lively, with dedicated venues (RSSRail, ICIRT) and even specialized formal tools for railways  (RobustRailS, SafeCap, OnTrack). 
Based on these findings, we highlight current research gaps and expected actions. In particular, the need to focus on more empirically sound research methods, such as Case Studies and Controlled Experiments, and to lower the degree of abstraction, by applying formal methods and tools to development phases that are closer to software development. Our study contributes with an empirically based perspective on the future of research and practice in formal methods applications for railways. It can be used by formal methods researchers to better focus their scientific inquiries, and by railway practitioners for an improved understanding of the interplay between formal methods and their specific application domain. 
%Our results on recent trends show that the field is lively, with dedicated venues (RSSRail, ICIRT) and even specialized formal tools for railways  (RobustRailS, SafeCap, OnTrack). 
%
%(vii) recent studies show that there is a higher specialization of the field, with dedicated tools (RobustRailS, SafeCap, OnTrack) and venues (RSSRail, ICIRT). Other \textit{hot} tools of recent years are UPPAAL (14\%) and ProB (13\%); (viii) industrial studies use semi-formal methods more frequently, and tend to give more importance to later phases of development. Some closed source tools have a clear industrial vocation (e.g., Prover Engines, S3 and IBM Rational), while other tools are applied almost exclusively in academic studies (e.g., CPN Tools and UPPAAL).
\end{abstract}

%\ccsdesc[300]{Software and its engineering~Correctness}
%
% End generated code
%


\maketitle

% The default list of authors is too long for headers.
\renewcommand{\shortauthors}{A. Ferrari and M. H. ter Beek}

\section{Introduction}
\label{sec:introduction}

%The railway signaling domain has traditionally been a fruitful playground for formal methods researchers. In 2004 Bjorner formulated the ``Grand Challenge'' of formal methods for railways~\cite{bjorner2004}, calling researchers and practitioners to a joint effort to formalise and share railway system models. The challenge is still ongoing in a broader sense, as railway systems are evolving, with increasing complexity due to higher automation and novel \hl{technological paradigms}~\cite{}. At the same time, the world of formal methods is changing, with the consolidation of novel techniques, such as \hl{Mau, here I need you}.

The railway signaling domain has traditionally been a fruitful playground for formal methods. The extensive survey of Woodcock \textit{et al.}~\cite{WLBF09} recognised transportation, including railways, as a primary field in which formal methods have been applied, also for the development of real-world railway platforms. Well-known  projects are Line~14 of the Paris Metro and the driverless Paris–Roissy Airport shuttle, developed with the B~method~\cite{Abr07}, the metro control system of Rio de Janeiro, developed with the support of Simulink/Stateflow~\cite{ferrari13metro}, and the verification of the ERTMS/ETCS European standard for railway control and management with NuSMV~\cite{chiappini2010}. 
A set of international joint projects has also been funded on formal methods for railways starting from 1998 (14 projects in total were counted until 2018, cf.~\cite{FBMBFGPT19}). Notable cases include OpenETCS (\url{http://openetcs.org}) and the more recent 4SECURail (\url{https://www.4securail.eu}) and X2Rail-1 (\url{https://cordis.europa.eu/project/id/730640}). Projects in the field have recently seen a particular boost also thanks to the Shift2Rail (\url{https://shift2rail.org/}) initiative. This is a joint effort of railway stakeholders and the European Commission to advance the railway field through innovative research projects involving academia and industry. Shift2Rail considers formal methods to be fundamental to the provision of safe and reliable technological advances in railways. 

Surveys on formal methods in industry, including railways, have already appeared in the literature. Some focus on providing personal overviews of past experiences~\cite{Abr07,Abr06,CW96} or on collecting viewpoints of experts~\cite{GM20,GBP20}; others target the railway field specifically, with questionnaires~\cite{BBFGMPTF18,BBFFGLM19}, discussion of future challenges~\cite{Fan13,Bou14}, and comparison of tools in the domain~\cite{Bjo03,ferrari2020comparing,mazzanti2018towards,FMBB21}.
However, despite the interest of the industry and research communities, there is no systematic study aimed at collecting and analyzing the existing literature in formal methods for railways to provide a framework to move forward in research and practice. This is particularly needed, as the world of formal methods is vast, and practitioners often face a paradox of choice in selecting formal techniques~\cite{ferrari2020comparing}. 

This paper presents the first systematic mapping study on formal methods in the railway domain. \rev{We focus on \textit{railway signaling}, given the long history of applications of formal methods in this field~\cite{Fan13}}. We retrieve and select 328 high-quality research papers from the literature in the time span 1989--2020, and we categorize them according to three different facets: \textit{demographic and empirical}, identifying years, publication venues and research methods used; \textit{formal methods}, categorizing techniques, tools and languages; \textit{railway}, concerning systems and development phases addressed by the research. Furthermore, we perform a stratified analysis to understand which are the characteristics of the studies concerned with industrial applications, and what are the main trends of the last years. 

Our results show that formal methods for railways is a thriving research field with a strong industrial bound, since 143~studies were published solely in the last five years~(44\% of the total), and 79~studies~(24\%) involve industry. Most of the studies focus on non-standard interlocking applications, and on high-level modeling and early development phases. In terms of languages and tools, the landscape is highly diversified. The dominant languages are UML~(18\%) and B~(15\%), while frequently used tools are ProB~(9\%), NuSMV~(8\%) and UPPAAL~(7\%), but a long tail exists in the statistics. The empirical maturity of the field is still limited, as many papers present only examples or experience reports. Our work thus calls for more empirical rigor in the field, with case studies, which can leverage the strong link with industries, and controlled experiments, which can address issues related to the learnability of formal methods and aspects related to human factors. Furthermore, we encourage applications that operate on later railway development phases, and on lower-level models and code, which received less attention so far. Finally, based on our findings, and in line with the needs for interoperability, we also support focusing more on modeling and verifying standard systems. 

The remainder of the paper is structured as follows. In Sect.~\ref{sec:motivation} we present background on formal methods and railways, and we discuss related reviews to motivate the current study. Sect.~\ref{sec:method} describes the review method. Sect.~\ref{sec:results} presents the results, and Sect.~\ref{sec:discussion} discusses the empirical findings. Sect.~\ref{sec:threats} reports threats to validity, and Sect.~\ref{sec:conclusion} concludes the paper.

\section{Background and Motivation}
\label{sec:motivation}

\subsection{Formal Techniques}
\label{sec:formaltechniques}

\emph{Formal methods\/} are rigorous mathematics-based techniques and tools for the specification (\emph{modeling\/}) and manual or automated verification (analysis) of software or hardware systems or system designs~\cite{WLBF09,GBP20}. 
\emph{Semi-formal methods\/} refer to techniques and tools that are not fully formal, i.e., lacking a precise and unambiguously defined syntax and semantics; a prominent example is the Unified Modeling Language (UML). 

\emph{Model-based development\/} relies on rigorous techniques (e.g., the B~family~\cite{BKKLLMV20}) to derive a concrete (low-level) implementation from an abstract (high-level) specification by successive \emph{refinement\/} steps based on \emph{model transformation\/}. During refinement, a specification is complemented with details that are unnecessary in the higher-level specifications. Model-based development usually involves some semi-formal methods and it is typically complemented with (automatic) \emph{code generation\/}, which generates source code that is by definition consistent with the model it is generated from.

\emph{Formal verification\/} concerns (exhaustive) verification that \rev{functional properties (e.g., absence of deadlocks) or} critical system properties related to safety and security are satisfied, i.e., verifying correctness of the system (model) which dynamic analysis methods based on \emph{simulation\/} or \emph{model-based testing\/} \rev{generally} cannot. Since neither of the latter two techniques explores all possible system behavior (state space), a counterexample found by either testing or simulation demonstrates an error but the lack of counterexamples does not prove absence of errors \rev{(e.g., deadlocks)}. The success of testing moreover depends on the quality of \emph{test generation\/}, which generates appropriate sequences of input values that guarantee the models to satisfy specific testing criteria (e.g., model coverage). \rev{Model-based testing has been applied successfully in the railway domain (e.g., using RT-Tester)~\cite{Pel13,BHHHPSH14,BFBT16,wang2018hybrid}.}

Formal verification is often (but not always) automated, and the resulting tools can reduce the effort and time needed to prove the correctness of systems considerably. Formal verification is supported by several families of techniques, the most important ones being \emph{theorem proving\/}~\cite{RV01} and \emph{model checking\/}~\cite{CHVB18}, including probabilistic~\cite{BK08} and statistical~\cite{AP18} approaches. 

Model checking verifies whether a system model meets a given specification, typically formulated in (temporal) logic; model checkers automate this process. Model checking explores all possible system behavior in the form of a (reachability) graph, possibly constructed on the fly. Model checking thus allows to determine the absence of errors in a system model \rev{and in case there is an error}, it moreover produces a counterexample that demonstrates how the error can be produced.
\emph{Reachability analysis\/} is used to determine which states of a model can be reached, thus suffering from the state-space explosion problem inherent to model checking: as the number of state variables of a system increases, the size of the system's state space grows exponentially. Advanced techniques such as \emph{symbolic\/} or \emph{bounded\/} model checking alleviate this problem, sometimes in combination with SAT/SMT solving~\cite{BHMW09}. 

Constraint satisfaction problems are decision problems stated in the form of a set of constraints that can be solved with \emph{SAT solving\/}, \emph{SMT solving\/} or (integer) \emph{linear programming solving\/} techniques~\cite{Tsa93}.
\emph{Static analysis\/} is another abstract interpretation technique to detect erroneous run-time behavior at compile-time, typically by computing over-approximations, i.e., including behavior that cannot actually occur. 
\emph{Static checking\/} is another compile-time technique, which catches syntax or typing errors, i.e., errors that are independent of specific variable values. 

Theorem proving uses deductive reasoning to provide a proof in symbolic logic by inference; theorem provers automate much of this process~\cite{NMLSI19}. Contrary to model checking, which is largely limited to finite models and propositional logic, theorem proving can handle infinite state spaces and many theorem provers moreover support automatic code generation.

Correct-by-construction approaches such as (supervisory control) \emph{synthesis\/}~\cite{CDLX02} concern the creation of (program or system) models that provably satisfy a high-level formal specification. In supervisory control synthesis, starting from a model of the uncontrolled system and a model of the behavioral requirements, a supervisory controller model is synthesized; a few tools to do so exist. Such a supervisory controller thus influences system behavior by disabling controllable events to guarantee system correctness with respect to the requirements (e.g., safety properties). 

\subsection{Railway Signaling Systems}
\label{sec:railwaysystems}

Railway signaling systems are complex, dependable cyber-physical platforms, composed of interacting subsystems with different safety-critical levels. These systems also have diverse applications, from traditional heavy rails, to light rapid rails and to metro lines. 

Signaling subsystems can be distinguished between those that mainly control the transit of trains at the stations, and those that mainly ensure safety along the lines. In a station, the most important subsystem is the so-called \textit{interlocking}, a safety critical platform that controls points and signals, and all the wayside entities, as, e.g., the elements to identify the presence of a train in a specific portion of the line (Axle Counters, or Track Circuits~\cite{antoni2011complementarity}). By monitoring and setting the status of the entities, the interlocking can enable safe train routing. In advanced metro systems, train routing is commanded by \rev{the so} called Automatic Train Supervision (ATS) platform, while in more traditional systems a human command is issued. 

Once a train is routed, preservation of the safety distance from other trains needs to be ensured. This is supported by Automatic Train Protection (ATP) platforms, which are composed of a wayside subsystem and an on-board subsystem. The wayside ATP monitors the position of the train, and makes preceding trains aware that the portion of the line in front of them is occupied. The on-board ATP receives information from the wayside one, and monitors the so-called braking curve, issuing an emergency brake in case there is a risk of collision.  Along the lines, a reduced form of interlocking, namely the Railway Crossing Controller, ensures the safety of level crossings. Advanced metro systems also include an automatic driver called Automatic Train Operation (ATO).

Given the need to ensure interoperability between the different subsystems described, product standards have been defined by international organizations. The most important product standard for heavy rail is the European Rail Traffic Management System/European Train Control System (ERTMS/ETCS)\footnote{\url{https://www.era.europa.eu/activities/european-rail-traffic-management-system-ertms\_en}}. The standard foresees four levels of automation (0 to 3), and from level 2 it includes the so-called Radio Block Center (RBC), a radio-based wayside ATP. The standard provides very detailed requirements specifications, as its goal is to ensure that subsystems developed by different vendors are able to seamlessly communicate, so to encourage competition. Another known standard for heavy rails is the Chinese Train Control System \rev{(CTCS)~\cite{lv2022chinese}}, which is analogous to the ERTMS/ETCS in terms of goals.  
A well-known product standard, oriented to metro lines, is the Communications-based Train Control (CBTC) system, also known as Urban Guided Transport Management and Command/Control System (UGTMS). Two international standards provide general requirements for CBTC systems, IEEE~1474.1-2004~\cite{ieee1474} and IEC~62290~\cite{iec62290}. The main characteristic of CBTC, shared also with ERTMS/ETCS Level 3, is the concept of \textit{moving block}. In a nutshell, this concept consists of computing the safety distance between trains considering the exact position each train, instead of considering as its position the segment of the line occupied by the train. The wayside ATP for CBTC systems is frequently called Zone Controller (ZC), though its name might depend on the vendor. 

The railway field is particularly lively in terms of innovation efforts, especially thanks to the increased sensitivity of the global community towards green transportation. At this regard, the Shift2Rail program (\url{https://shift2rail.org/}) is an unprecedented joint effort by the European rail sector and the European Union (EU), tripling EU-funding to nearly €1 Billion for rail research, innovation, and demonstration across the 7-year lifespan of the initiative, to move European Railway forward. In November 2015, the Shift2Rail Multi-Annual Action Plan was adopted and in June 2016 Shift2Rail awarded the first grants. Over 100 projects have been financed so far, including projects in which formal methods play a prominent role (ASTRail, X2Rail-1, X2Rail-2, 4SecuRail, \textit{etc.}\footnote{\url{https://projects.shift2rail.org/s2r_projects.aspx}}). A successor to the Shift2Rail joint undertaking, Europe's Rail, has recently been announced\footnote{\url{https://ec.europa.eu/commission/presscorner/detail/en/ip_21_702}}. \rev{Other notable initiatives outside the EU are the UK Rail Research and Innovation Network (UKRRIN)\footnote{\url{https://www.ukrrin.org.uk/}} and the Chinese State Key Laboratory of Rail Traffic Control and Safety\footnote{\url{http://en.bjtu.edu.cn/research/institute/laboratory/16583.htm}}.}

%Requirements from the standards are not detailed, as vendors normally develop entire CBTC systems, including ATP, ATO, ATS, etc., as metros are closed systems in which interoperability concerns are less prominent. 

\subsection{Related Reviews}
\label{sec:relatedreviews}

Formal methods have been studied in academia and applied in industry for quite some time now, as witnessed by introductions from the early '90s to the use of formal methods in developing safety-critical software systems from academia~\cite{Win90,PA93,Rus93} and industry~\cite{Tho90,Tho93}. 
Further historical references from long-time advocates of formal methods reflect on the industrial application of formal methods through the metaphors of seven (and seven more) myths~\cite{Hal90,BH95b} and ten commandments~\cite{BH95a,BH06} to eventually realize their benefits~\cite{Hal07}.
Also worth mentioning are early experiences and perspectives on industrial applications of formal methods~\cite{BS93,GCR94}, as well as the first systematic survey and analysis of the use of formal methods in the development of industrial applications~\cite{CGR95a}, publicized by several works~\cite{CGR92,CGR95b,Cra95}. This extensive survey is based on twelve case studies from industry, including one from the railway signaling domain, namely the development of ATP systems for the subways of Paris and Calcutta~\cite{GH90,DDM92}.
The mid-'90s were also characterized by panels and round-table discussions, involving academics and practitioners alike, on the (future) use of formal methods in industry~\cite{HPPRS95,BBDGGHHHJJLPRWZ96}. 

The classical 1996 survey on formal methods~\cite{CW96} illustrates a number of case studies in specification and verification, including the one from the railway signaling domain described in~\cite{CGR95a} plus an additional one on specifying the signaling rules of railway interlocking systems~\cite{Kin94}.

At the turn of the century, several personal, non-systematic surveys, often based on previous surveys and involving practitioners,
%and mainly from practitioners or aimed at practitioners, 
were published~\cite{Cra99,HB99,BCKUW00}; Dietrich and Hubaux~\cite{DH02} present a more extensive survey of the use of formal methods for communication services 
%(not protocols) 
both in academia and in industry. 
Shortly after, the first non-systematic surveys of formal methods in the railway domain were published~\cite{Bjo03,PB04}; these are both very personal, informal reviews of formal techniques and tools and exemplary applications to railway systems.
Also worth mentioning are a tutorial introduction to the B~method~\cite{Abr07} and a brief description and discussion of two of its best-known applications in industry~\cite{Abr06}: the development of safety-critical parts of the subway line~14 and the Roissy airport shuttle of Paris~\cite{BBFM99,BA05}.

The classical 2009 survey on formal methods~\cite{WLBF09}, \lq\lq perhaps the most comprehensive review ever made of formal methods application in industry\rq\rq, reviews the application of formal methods in 62~different industrial projects world-wide, in all but 6~cases by collecting data directly from individuals who had been involved in the projects. One of the eight highlighted projects is from the railway domain. The paper also provides an overview of 20 years of surveys on formal methods in industry, including all surveys mentioned above. 

The last decade has seen several introductions and accounts of trends and experiences concerning the role of formal methods in the development of safety-critical applications~\cite{Hax10,GM13,BH14}, in particular in the railway domain and from an academic viewpoint by Fantechi et al.~\cite{Fan12,Fan13,FFM13,FFGM13} and in the general transportation domain and from an industrial viewpoint, focusing mostly on SCADE and/or the B~method, by Boulanger~\cite{Bou12a,Bou12b,Bou14}. 
Further studies are lessons learned and obstacles found with respect to decades of integrating formal methods in research, education and industrial practice, in particular in the transportation domain~\cite{Mil12,DCCFHHHMW13,BjH14,LDPM17,BKKLLMV20}. 
We also mention a number of recent surveys in the railway domain at large~\cite{HOWM15,FYY15,MRH015,XZAZ16,UPSSJA16,MBB17,LLB18}, neither of which involve formal methods.
The extensive report from Garavel and Graf~\cite{GG13} provides a state-of-the-art account on the use of formal methods in academia and industry, and a large number of success stories: a carefully selected list of 30, well-documented case studies from three decades (one per year from the period 1982--2011), including a large number of railway cases. 
%all those from the railway domain mentioned above.
Finally, we mention three recent questionnaire surveys on the use of formal methods in the railway domain~\cite{BBFGMPTF18,FBMBFGPT19,BBFFGLM19}, conducted with both academic and industrial stakeholders in the context of Shift2Rail, %(\url{https://shift2rail.org/}), a joint undertaking by the European rail sector and the European Union, 
which identifies the main formal methods and tools used in the railway domain as well as their most relevant functionalities and features.

The recent 2020 survey on formal methods~\cite{GBP20}, \lq\lq an unprecedented effort to gather the collective knowledge of the formal methods community\rq\rq, reviews the responses of 130 high-profile experts in formal methods to 30 questions on the past, present and future of formal methods in research, industry and education. The paper also presents 111~position statements by these experts about the challenges and benefits of formal methods. 
In parallel, \cite{GM20}, \lq\lq the largest cross-sectional survey of formal methods use among software engineering researchers and practitioners to this date\rq\rq, surveys the academic and industrial use of formal methods in safety-critical software domains, identifying transportation as a typical application domain for formal methods.
%without providing any particular attention to the railway domain.    

\paragraph{Contribution} The evidence from previous reviews shows the interest of the research community in formal methods for railways, as well as several notable attempts of surveying literature and practitioners in the field. 
%At the same time, the Shift2Rail program  witnesses the interest of the \textit{civil} community in railway advancements. 
Nevertheless, the vast majority of the studies are either non-systematic, based on personal opinions, or collecting information from stakeholders or experts rather than from scientific literature. To our knowledge, this is the first systematic mapping study on the topic. \rev{The only study that can be compared to ours is the workshop paper by Gruner \textit{et al.}~\cite{gruner2015towards}, who aimed at identifying settled knowledge in the railway domain. This work is more preliminary, as it considers only three reference conferences as data sources, but broader in terms of scope, as it aims to cover the entire railway domain.} \rev{We remark that our study focuses on the prominent field of railway signaling systems~\cite{Fan13}, while we are not concerned with other railway applications, such as, e.g., rolling stock or passenger handling. In the following, when we use the term ``railway'' we generally mean ``railway signaling''.} 

Our work contributes to the body of knowledge with an empirically grounded overview of the current state of the research on formal methods in railways, to help researchers and practitioners identify the current gaps, and embrace future challenges. 

\newpage
\section{Review Method}
\label{sec:method}
\rev{The current survey is a systematic mapping study (SMS), which can be regarded as a variant of systematic literature reviews aiming at \textit{classifying} the literature, rather than synthesising evidence~\cite{kitchenham2004procedures}.}   
This section describes the review method adopted, which follows the guidelines of Kitchenham~\cite{kitchenham2004procedures} for conducting literature reviews, \rev{and which are largely used also for SMSs~\cite{petersen2015guidelines}.}  
Accordingly, we first outline research questions (Sect.~\ref{sec:questions}), and then we illustrate the search string (Sect.~\ref{sec:string}), the study search and selection strategy (Sect.~\ref{sec:studyselectionprocedure}), followed by data extraction (Sect.~\ref{sec:extraction}) and data synthesis procedures (Sect.~\ref{sec:synthesis}). 

\subsection{Review Questions}
\label{sec:questions}

The main goal of our \rev{SMS} is as follows:

\smallskip
\textbf{Goal:} \textit{Identify, understand and characterize studies on the application of formal methods to the development of railway systems, identifying recent trends and considering industrial applications, for the purpose of supporting formal methods practice and research.}
\smallskip

To address this goal, we aim to first provide a demographic characterization of studies concerning applications of formal methods in the railway domain. Then, we plan to classify which methods are used, in which phase of the development process, and for what types of systems.  Within this classification, we want to identify which papers are concerned with industrial applications, either in the context of exploratory studies involving industrial partners, or for the development of real-world railway products. Finally, we want to focus on the trends of the last years to understand possible future directions. By \lq recent\rq\ we intend studies published after 2015, as in our research we identified a relevant increase in the number of studies starting from 2016 (cf.\ Fig.~\ref{fig:year}). Therefore, from our goal, we derive the following research questions. 

\begin{itemize}
\item RQ1: How is research demographically and empirically characterized in the field of applications of formal methods in the railway domain? 
\begin{itemize}
\item RQ1.1: What is the \textbf{time} distribution of primary studies?
\item RQ1.2: What is the \textbf{venue} distribution of primary studies? 
\item RQ1.3: Which type of \textbf{evaluation} has been conducted in the primary studies? %about applications of formal methods in the railway domain?
\item RQ1-I: What is the degree of \textbf{industrial} involvement in the primary studies?
\end{itemize}
%\item RQ2: What primary studies have been conducted about applications of formal methods in the railway domain?
%\begin{itemize}
\item RQ2: What formal methods are used in the railway domain? 
\begin{itemize}
\item RQ2.1: What is the degree of \textbf{formality} of the studies?
\item RQ2.2: What \textbf{formal techniques} are used?
\item RQ2.3: Which specification \textbf{languages}?
\item RQ2.4: Which \textbf{tools}?
\end{itemize}
\item RQ3: In which way are formal methods applied to railway system development? 
\begin{itemize}
\item RQ3.1: To which category of \textbf{railway system}? 
\item RQ3.2: To which category of \textbf{railway subsystem}?
\item RQ3.3: In which \textbf{phases} of the system development are formal methods applied?
\end{itemize}
%\item RQ2.3: In which phases of the system development are the formal methods applied?
%\item RQ2.4: To which type of railway systems are the formal methods applied?
%\end{itemize}
\item RQ-I: What are the characteristics of the studies reporting \textbf{industrial} applications?
\item RQ-T: What are the emerging \textbf{trends} of the last years?
\end{itemize}

RQ1 aims to give a first overview of the time and venue distribution of the studies, to help identifying the evolution of the field across time, relevant journals and conferences, and the empirical maturity of the studies. RQ1 also includes a ``service'' question, namely RQ1-I, which serves to support the stratified analysis in relation to RQ-I. 
RQ2 focuses on the formal methods facet, identifying the core elements of any formal method, namely degree of formality, technique, language and tool. 
RQ3, instead, focuses on the railway facet, and aims to identify the most common phases, railway systems and subsystems in which formal methods are applied. 

RQ-I and RQ-T, instead, aim to address \textit{recent trends and industrial applications}, as specified in our overarching research goal. While RQ1 to RQ3 are independent questions, RQ-I and RQ-T are cross-cutting questions (e.g., certain methods identified in RQ2 may be more trendy, other more established and industrially validated). The paper is organized to primarily answer RQ1, RQ2, and RQ3 while RQ-T and RQ-I are answered in relation to the other questions to facilitate the interpretation of the data and have a more concise visualization of the statistics. 

\subsection{Databases and Search String}
\label{sec:string}
We selected the following scientific databases as data sources, which typically include papers in our considered scope: ACM Digital Library, IEEE Xplore, ScienceDirect and SpringerLink. For SpringerLink, we focus the search on the categories of ``Computer Science $>$ Software Engineering'' and ``Engineering $>$ Software Engineering/Programming and Operating Systems'', as a pilot search on the entire database---which supports full-text search only--- conducted to 49,116 documents, a number that was considered unmanageable for the available resources. 
To define the search string, we use the major terms ``formal methods'' (representing the object of the research, or intervention) and ``railways'' (representing the domain of application, or context) as base terms. Then, we elaborate each base term with alternative words, keyphrases and wildcards, when appropriate. We then use the Boolean OR to incorporate alternatives into each base term set, and Boolean AND to link the two sets. The terms were initially selected based on brainstorming among the participants and the search string was further refined through pilot searches. The final search string:\\

\begin{centering}
``formal'' OR  ``model check*'' OR ``model based'' OR  ``model driven'' OR \\  ``theorem prov*'' OR ``static analysis'' \\
\textbf{AND} \\
``railway*'' OR  ``CBTC'' OR   ``ERTMS'' OR ``ETCS'' OR ``interlocking'' OR  \\  ``automatic train'' OR     ``train control'' OR ``metro'' OR    ``CENELEC''\\
\end{centering}

%The Pilot Test enabled to assess that, among the 40 studies considered, 24 of them resulted to be relevant for the RQs. This leads to a precision of 60\%, which we consider acceptable for our goal. Recall cannot be evaluated on the Pilot Test. Identification of relevant studies possibly not identified by means of the secondary search strategy.

\begin{table}[]
\begin{tabular}{|c|p{0.925\textwidth}|}
\hline
\multicolumn{2}{|c|}{\textbf{Inclusion Criteria}}\\ \hline
I1 & The study presents an application of formal or semi-formal method, including model-based development methods, to the development of railway systems \\ \hline
I2 & The study is mainly concerned with the development of railway systems for signaling and control\\ \hline
I3 & The study comes from an acceptable source such as a peer-reviewed scientific journal, conference, symposium, or workshop\\ \hline
I4 & The study is written in English language\\ \hline
\hline
\multicolumn{2}{|c|}{\textbf{Exclusion Criteria}}\\ \hline 
E1 & The study does not use a formal or semi-formal method \\ \hline
E2 & The study does not apply a method to the railway domain \\ \hline
E3 & The study uses a railway problem as a part of a benchmark for performance evaluation \\ \hline
E4 & The study is concerned with the quantitative assessment of reliability, availability and maintainability (RAM) requirements expressed in quantitative form \\ \hline
E5 & The type of study is a secondary study \\ \hline
E6 & The type of study is a book or a book section \\ \hline
E7 & The study did not undergo a peer-review process (i.e., Festschrift contributions, etc.) \\ \hline
E8 & The study has been published in another, extended form \\ \hline
E9 & The study has the form of editorial, abstract, keynote, poster or a short paper (body less than or equal to 4 pages) \\ \hline
\end{tabular}
\caption{\label{tab:inclusionexclusioncriteria}Inclusion and exclusion criteria.}
%\end{table}
%
%\begin{table}[]
\begin{tabular}{|r|p{0.925\textwidth}|}
\hline
1 & Is there a clear statement of the aims of the research?\\ \hline
2 & Is there an adequate description of the context in which the research was carried out?\\ \hline
3 & Is there a clear description of the adopted research methodology?\\ \hline
4 & Is there a clear description of the task addressed with formal methods? \\ \hline
5 & Is there a clear identification of the formal languages and tools used? \\ \hline
6 & Was the data collected in a way that addressed the research issue?\\ \hline
7 & Was the data analysis sufficiently rigorous?\\ \hline
8 & Is there a clear statement of the findings and limitations?\\ \hline
9 & Is the study of value for research?\\ \hline
10 & Is the study of value for practice?\\ \hline
\end{tabular}
\caption{\label{tab:qualitycriteria}Quality checklist adopted for the study.}
\end{table}

\subsection{Search Strategy and Study Selection Procedures}
\label{sec:studyselectionprocedure}

Our \textbf{primary search} strategy consists of adapting the search string to each specific database, and then selecting relevant studies for data extraction. The selection is performed considering inclusion and exclusion criteria listed in Table~\ref{tab:inclusionexclusioncriteria} and the quality checklist listed in Table~\ref{tab:qualitycriteria}. \rev{Inclusion and exclusion criteria were developed during pilot searches, to restrict the scope to comparable studies in railway signaling and control (I1--I2, E1--E4) and ensure quality and representativeness (I3--I4, E5--E9).}
The quality checklist is inspired by the work of Dyb{\aa} and Dings{\o}yr~\cite{dyba2009empirical} and Chen and Ali Babar~\cite{chen2011systematic}, and adapted to our context during initial pilots. The selection procedure is carried out according to the following instructions:

\begin{enumerate}
    \item \textbf{Retrieval:} adapt the search to each specific search engine, considering its peculiarities. Perform the search on metadata only. If this is not supported, perform full-text search. 
    \item \textbf{Screening:} read title and abstract of the papers and apply the inclusion criteria. The papers that fulfill the criteria are marked as \textit{included}, they are downloaded and stored in Zotero. Use the features of Zotero to identify and discard duplicates. 
    \item \textbf{Full-text Reading:} read the full-text of the included papers, and apply the exclusion criteria, plus the quality checklist. A ternary scale ($\text{Yes} = 1$, $\text{Partial} = 0.5$, $\text{No} = 0$) is used to grade the studies on each question reported in the checklist. The quality score of the paper is computed as the sum of the grades. If a paper does not not reach a quality score higher than~6 out of~10, exclude the paper from the selection. This threshold was identified during a pilot study to discard papers that would not allow appropriate data extraction, as relevant information is missing. The selected papers are given a unique identifier and are retained for data extraction. 
\end{enumerate}

The \textbf{secondary search} strategy consists of identifying additional studies by performing backward snowballing on high-quality papers, namely all those papers that received a quality score equal to~10---all quality criteria fulfilled. 

Two researchers---first and second author---apply the study selection procedures outlined above on separate subsets of the retrieved studies. When a study includes the researcher among the authors, the selection is performed by the other researcher to reduce bias. The two researchers have backgrounds in formal methods and railways, but complementary competences. The first author has a more industrial background, having worked as system engineer in a railway company, and participated to several technology transfer studies. The second author has a strong academic background on different formal methods, formal languages and logic. Thus, possible deficiencies in terms of knowledge of formal methods of the first author are compensated by the second one, while knowledge gaps on railway-specific aspects by the second author are addressed by the first one. These considerations apply also to the data extraction and synthesis procedures (Sect.~\ref{sec:extraction}). 

\subsubsection{Performing Primary Search and Study Selection}

The primary search is repeated four times on the following dates: October 30, 2017 (Pilot); December 7, 2017 (First, I); November 26, 2018 (Second, II); September 24, 2020 (Third, III). 

\begin{figure}[h]
\centering
\includegraphics[width=1\textwidth]{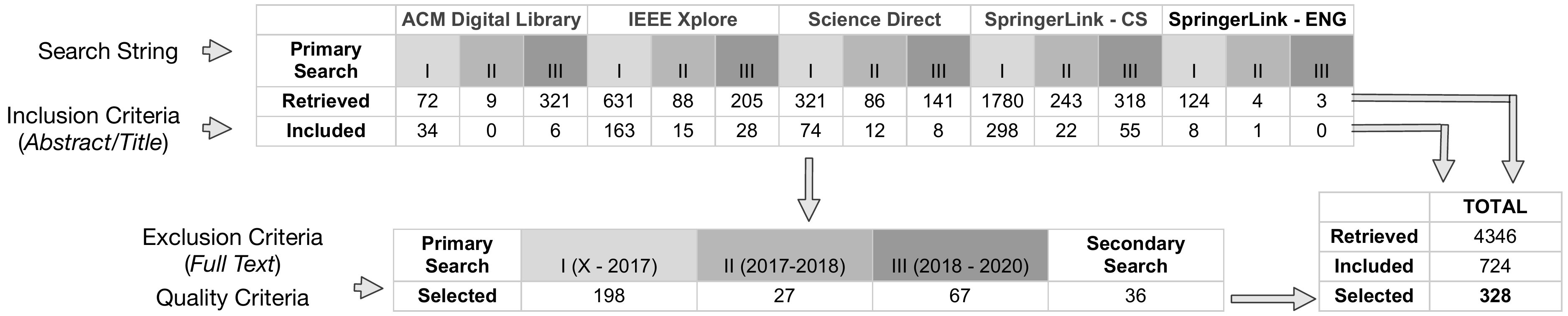}
\caption{\label{fig:process}Process of study selection and numerical results. \rev{The value of TOTAL in the bottom-right table is obtained by summing-up the cells in Retrieved, Included and Selected from the other tables.}}
\end{figure}

The search of October 2017 was conducted as part of a pilot study, in which also the data extraction criteria were piloted on a sample of the top-10 papers retrieved by the four search engines (40~papers in total). The pilot study allowed the involved assessors to align their judgments in data selection and extraction, and consolidate the procedures. Since the search, selection and analysis were focused on a subset of the papers, we do not report these results here. 

Search I was unbounded, so all papers available before December 7, 2017 were considered. The other searches were restricted to the time interval between the last year of the preceding search, and the year of the search\footnote{This choice prevented from ignoring papers at the boundary between years. Duplicates were discarded with the support of Zotero (cf.\ the study selection procedure in Sect.~\ref{sec:studyselectionprocedure})}. Therefore, Search II collected papers in the interval Jan~1, 2017---Nov~26, 2018, and Search III in Jan~1, 2018---Sep~24, 2020. 

The secondary search was conducted based on the papers selected in the primary search. Fig.~\ref{fig:process} reports the numerical results of the study selection procedure. Overall, 4,346~studies were initially retrieved, 724~were included, and \textbf{328} were finally selected.

% Resources to be searched are:
% \begin{itemize}
% \item ACM Digital Library (Pilot: 71)
% \item IEEE Explore Digital Library (Pilot: 624)
% \item SpringerLink (Pilot: 49,116)
% \item ScienceDirect (Pilot: 307)
% \end{itemize}

\begin{table}[h]

\begin{tabular}{|p{0.25\textwidth}|p{0.7\textwidth}|}

\hline
\multicolumn{2}{|c|}{\textbf{Empirical Evaluation Information (RQ1.3)}}                                                                                                                                                                                                                                                                                                                  \\ \hline
\multirow{11}{0.25\textwidth}{\textbf{Type of Study and Evaluation (from~Chen and Ali~Babar~\cite{chen2011systematic})}}  & Rigorous analysis (RA): Rigorous derivation and proof, suited for   formal model                                                                                                                                                                              \\ \cline{2-2} 
                                                                                  & \textbf{Case study (CS):} An empirical   inquiry that investigates a contemporary phenomenon within its real-life   context; when the boundaries between phenomenon and context are not clearly   evident; and in which multiple sources of evidence are used \\ \cline{2-2} 
                                                                                  & \textbf{Discussion (DC):} Provided some   qualitative, textual, opinion                                                                                                                                                                                       \\ \cline{2-2} 
                                                                                  & \textbf{Example (EX):} Authors describing an   application and provide an example to assist in the description, but the   example is used to \lq validate\rq\ or \lq evaluate\rq\ as far as the authors suggest                                                          \\ \cline{2-2} 
                                                                                  & \textbf{Experience Report (ER):} The result   has been used on real examples, but not in the form of case studies or   controlled experiments, the evidence of its use is collected informally or   formally                                                 \\ \cline{2-2} 
                                                                                  & Field study (FS): Controlled experiment performed in industry settings                                                                                                                                                                                        \\ \cline{2-2} 
                                                                                  & Laboratory experiment with human subjects (LH):   Identification of precise relationships between variables in a designed   controlled environment using human subjects and quantitative techniques                                                           \\ \cline{2-2} 
                                                                                  & Laboratory experiment with software subjects (LS): A   laboratory experiment to compare the performance of newly proposed system   with other existing systems                                                                                                \\ \cline{2-2} 
                                                                                  & Simulation (SI): Execution of a system with artificial   data, using a model of the real world                                                                                                                                                                 \\ \hline
                          \end{tabular}

\caption{\label{tab:evaluationinfo}Data extraction categories for evaluation information, which are used to answer RQ1.3. The types of study highlighted in bold are the ones that have actually been found in the selected papers.}
%\end{table}  
%
%\begin{table}[h]
%
\begin{tabular}{|p{0.25\textwidth}|p{0.7\textwidth}|}

\hline
\multicolumn{2}{|c|}{\textbf{Industrial Study Information (RQ1-I)}}                                                                                                                                                                                                                                                                                                                  \\ \hline                                                                                  
\multirow{4}{0.25\textwidth}{\textbf{Industrial Evaluation (adapted from Chen and Ali Babar~\cite{chen2011systematic})}} & NO: not evaluated in industrial settings                                                                                                                                                                                                                      \\ \cline{2-2} 
                                                                                  & LAB: industrial problem treated in laboratory settings                                                                                                                                                                                                        \\ \cline{2-2} 
                                                                                  & IND: industrial problem validated with railway experts                                                                                                                                                                                                        \\ \cline{2-2} 
                                                                                  & DEV: development of an industrial product                                                                                                                                                                                                                     \\ \hline
\multirow{3}{0.25\textwidth}{\textbf{Authorship}}                                              & A: only academic authors                                                                                                                                                                                                                                      \\ \cline{2-2} 
                                                                                  & I: only industrial authors                                                                                                                                                                                                                                    \\ \cline{2-2} 
                                                                                  & AI: both academic and industrial authors                                                                                                                                                                                                                           \\ \hline
\end{tabular}

\caption{\label{tab:industrialevaluation}Data extraction categories used to identify industrial studies, in relation to RQ1-I.}
\end{table}

\begin{table}[h]

\begin{tabular}{|p{0.25\textwidth}|p{0.7\textwidth}|}

\hline
\multicolumn{2}{|c|}{\textbf{Formal Methods Facet (RQ2)}}                                                                                                                                                                                                    \\ \hline
                                                                                       & F: Formal                                                                                                                                                                                       \\ \cline{2-2} 
                                                                                       & SF: Semi-formal                                                                                                                                                                                 \\ \cline{2-2} 
\multirow{-3}{0.25\textwidth}{\textbf{Degree of formality of the method(s)}}                        & SFF: Semi-formal and Formal                                                                                                                                                                     \\ \hline
\textbf{Name(s) of the technique(s) applied}                                           & List of techniques applied in the paper (e.g., model-based development, model checking, theorem proving)                                                                                        \\ \hline
\textbf{Name(s) of the language(s) used}                                               & Name of the languages used for modeling in the context of the paper (e.g., UML, State Machines)                                                                                                \\ \hline
\textbf{Name(s) of the support tool(s) used}                                           & Name of the tools used in the paper (e.g., Atelier B, SCADE, Simulink, UPPAAL)                                                                                                                  \\ \hline
\end{tabular}
\caption{\label{tab:fmcontextdescription}Data extraction categories related to the formal methods facet, used to answer RQ2.}
%\end{table}
%
%
%
%\begin{table}[h]
\begin{tabular}{|p{0.25\textwidth}|p{0.7\textwidth}|}
\hline
\multicolumn{2}{|c|}{\textbf{Railway Context (RQ3)}}                                                                                                                                                                                                   
\\ \hline
                                                                                       & P: Planning                                                                                                                                                                                     \\ \cline{2-2} 
                                                                                       & R: Requirements                                                                                                                                                                                 \\ \cline{2-2} 
                                                                                       & A: Architecture \& Design                                                                                                                                                                               \\ \cline{2-2} 
                                                                                       & D: Detailed Design                                                                                                                                                                             \\ \cline{2-2} 
                                                                                       & I: Implementation                                                                                                                                                                               \\ \cline{2-2} 
                                                                                       & T: Testing                                                                                                                                                                                      \\ \cline{2-2} 
                                                                                       & N: Integration                                                                                                                                                                                  \\ \cline{2-2} 
                                                                                       & V: Validation                                                                                                                                                                                   \\ \cline{2-2} 
\multirow{-9}{0.25\textwidth}{\textbf{Phase(s) of the system development addressed (from~CENELEC)}} & M: Maintenance                                                                                                                                                                                \\ \hline
                                                                                       & SA (Stand-alone system): if the system treated in the paper is not related with other systems, as for the case of Platform Screen Door Controllers, Railway Crossing Controllers, Axle Counters \\ \cline{2-2} 
                                                                                       & ERTMS-ETCS: if the system is part of an ERTMS/ETCS system                                                                                                                                       \\ \cline{2-2} 
                                                                                       & CBTC: if the system is part of a CBTC system                                                                                                                                                    \\ \cline{2-2} 
                                                                                       & CTCS: if the system is part of a CTCS system                                                                                                                                                    \\ \cline{2-2} 
\multirow{-5}{0.25\textwidth}{\textbf{Type(s) of railway system considered}}                        & NS (Non-standard Train Control and Management): if the system is part of a train control and management system that does not follow an international standard                                   \\ \hline
                                                                                       & For Stand-alone systems (SA): name of the system (e.g., Platform Screen Door, Railway Crossing, Axle Counter, etc.)                                                                           \\ \cline{2-2}   
\multirow{4}{0.25\textwidth}{\textbf{Type(s) of railway subsystem considered}}                      & For ERTMS-ETCS, CBTC, CTCS, NS:                                         
                                                            \begin{itemize}[leftmargin=*,nosep,after=\vspace{-\baselineskip}]
                                                                \item HLCL (High-level Control Logic and Communication): if the paper treats the high-level logic of the system, or the communication between two or more components 
                                                                \item Name of the system (e.g., wayside ATP, onboard ATP, ATO, Radio Block Center, etc.): if the paper treats solely one subsystem
                                                            \end{itemize}
\\ \hline 
\end{tabular}

\caption{\label{tab:railwaycontextdescription}Data extraction categories in relation to the railway context, used to answers RQ3.}
\end{table}

% Please add the following required packages to your document preamble:
% \usepackage{multirow}

\subsection{Data Extraction Procedure}
\label{sec:extraction}

Data extraction is performed by the authors, referred in the following as \textit{extractors}. Relevant publications are partitioned into two balanced sets, and each extractor extracts data from one set. Each extractor reviews also the extracted data for the other set, \rev{after reading the associated papers}. In case of disagreement, a third expert in railway applications to railway problems, namely Alessandro Fantechi from the University of Florence\footnote{The author is the most cited in Scopus when searching for publications in ``formal method'' and ``railway''.}, is consulted to guide towards a decision on the data to be recorded. When one of the extractor\rev{s} is author of a relevant study, the data extraction is conducted by the other. The data is recorded in the form of shared Google spreadsheets to ease analysis and cross-checking.

Data extraction is conducted by first extracting publication details---title, authors, type of venue (conference or journal), name of venue, publication year and doi. This information is used to answer RQ1.1 and RQ1.2. In addition, the information about the year of publication serves also RQ-T. To answer RQ.1.3, we extract evaluation information following the categorisation proposed by Chen and Ali Babar~\cite{chen2011systematic}, reported in Table~\ref{tab:evaluationinfo}. Furthermore, to answer RQ1-I we extract the information from Table~\ref{tab:industrialevaluation}, also adapted from Chen and Ali Babar~\cite{chen2011systematic}, and enriched with information about authorship. This information serves also RQ-I, \rev{besides RQ1-I}. 
Concerning the subquestions of RQ2, about the formal method facet, we extract the data reported in the extraction scheme of Table~\ref{tab:fmcontextdescription}, while for RQ3 the scheme in Table~\ref{tab:railwaycontextdescription} is applied. The schemes reported in the tables include two types of data: those for which pre-defined classes are considered (reported in \textbf{bold} in the tables); those for which the extractors can use free-text (reported in \textbf{\textit{bold italic}}). The free text is homogenized in the data synthesis procedure. The extraction schemes also allow the possibility to include more than one element in each extraction item, e.g., more than one language or more than one phase. \rev{The extraction schemes outlined in this section are the \textit{classification schemes} of our SMS.}

\subsection{Data Synthesis Procedure}
\label{sec:synthesis}

In the data synthesis procedure we consider the extracted data, we homogenize them and provide visual analytics to systematically answer the RQs. Part of the data (e.g., all evaluation information, as well as the degree of formality of the method, phases, type of study) are well-defined sets of classes. For these cases, data synthesis is straightforward, and results are represented in the form of graphical diagrams, choosing the most appropriate for each case to ease visualization and analysis. Other data (e.g., name of tools, languages, techniques) are based on free-text entered by the extractors. 
%As mentioned, free-text types of data are reported in \textit{bold italic} in the tables. 
This data needs to be homogenized, and, to this end, we adopt an open coding technique~\cite{saldana2021coding}. One author codes the free text, and produces a set of well-defined and finite set of tags that can be used to produce appropriate statistics and data visualization. The other author reviews and cross-checks the coding results also in relation to understandability and clarity of the tags. In this task, the first author primarily homogenizes data in relation to RQ3, given his greater expertise in railway systems. The second author, instead, primarily homogenizes data in relation to RQ2, given his broader knowledge of formal methods. This process is carried out throughout multiple iteration until an agreed set of classes is reached also for the free-text fields.

In general, we visually synthesize results by strictly following the categories identified, and by providing histograms that account for industrial studies and for recent trends. In some cases we considered it appropriate to provide refined data synthesis, guided by evidence from the extracted data. Specifically, the studies were observed to use multiple formal techniques, and thus we synthesize data also about combinations of techniques. Similarly, for railway development phases, we highlight combinations of multiple phases addressed by the same study. Concerning tools, we provide combined statistics with specification languages, to highlight relevant relationships, as we observed that often the adopted specification language does not match with the used tool. Finally, relationships between categories of systems and subsystems are also highlighted. 

The final spreadsheet file, which has been used to produce the statistics in this paper, is publicly shared at  \url{https://doi.org/10.5281/zenodo.5084640}.
\section{Results}
\label{sec:results}
In the following, we report the results of our analysis. 
%We first present statistics in relation to the cross-cutting RQs, concerned with industrial applications (RQ-I) and recent trends (RQ-T), as these are useful to perform a stratified analysis for the rest of the RQs. 
%Then, we report the answer to the RQs.
Specifically, for each element of interest, we first plot and discuss the total number of studies in relation to industrial ones. Then, we plot the distribution for recent studies (published after 2015) without distinguishing between industrial and academic. We observed that industrial studies are scarce in recent years (cf.\ Sect.~\ref{sec:year}), thereby limiting the relevance of associated statistics, which will not be reported. 

To support the reporting, we first answer the \lq service\rq\ question RQ1-I, which allows us to perform a stratified analysis in relation to RQ-I. Then follows RQ1.1, to support stratified analysis for RQ-T. 

\begin{figure}[h]
\centering
\includegraphics[width=.85\textwidth]{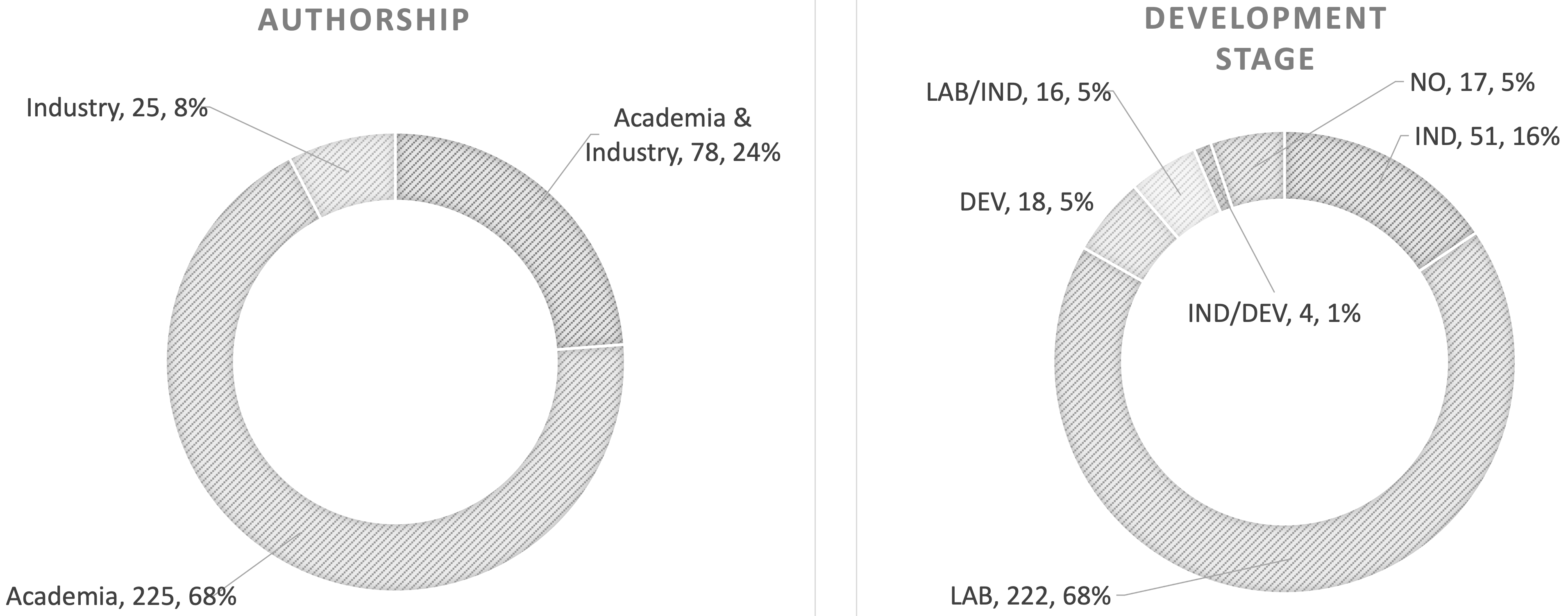}
\caption{Degree of industrial involvement, evaluated in terms of authorship of the paper and development stage of the study.}
\label{fig:authorship}
\end{figure}

\subsection{RQ1-I: Industrial Involvement}
\label{sec:industrialapp}

Fig.~\ref{fig:authorship} reports the degree of industrial involvement identified in the studies, based on the type of authorship and the development stage of the application considered in the paper. 

We see that about two thirds of the studies~(222, 68\%) have academic authors only, while the other third have some form of industrial involvement, either in conjunction with academic authors~(78, 24\%) or with authors coming exclusively from industry~(25, 8\%). 

Looking at the development stage we see a similar situation, with the majority of the studies concerned with industrial problems treated in laboratory settings~(LAB, 222, 68\%). In a non-negligible amount of cases, however, a step forward was performed: part of the studies have been validated with railway experts~(IND, 51, 16\%), part of them document the development of railway products with formal methods~(DEV, 18, 5\%) and some studies were considered  somewhat in between the different categories~(LAB/IND, 16, 5\%; IND/DEV, 4, 1\%). Still, some studies did not have any form of contact with industry and its specific problems, but considered solely toy problems~(NO, 17, 5\%).  

The statistics just described are used in the following to identify those studies that are concerned with industrial applications (industrial studies, for short)  and perform a stratified analysis of the results. Specifically, we consider a study to be \textit{industrial} in two main cases: (i)~it is tagged as IND, DEV, or IND/DEV; (ii)~it is tagged as LAB/IND and the authorship is AI or~I. In this way, we exclude from industrial studies borderline cases for which the actual industrial involvement is not entirely clear (i.e., those marked as LAB/IND, but without industrial authors). According to this classification, 24\% of the studies are industrial (79~in total), while~76\% are not~(249)---in the following we refer to these studies as \textit{academic}.  

The results show that studies are in general performed by academics in laboratory settings. Nevertheless, a relevant number of them (about one fourth) involve industry in some form, with formal methods applied also for the development of real products. This proportion suggests that research in the field is not self-referential and interaction with industry is present. 

\begin{figure}[h]
\centering
\includegraphics[width=1\textwidth]{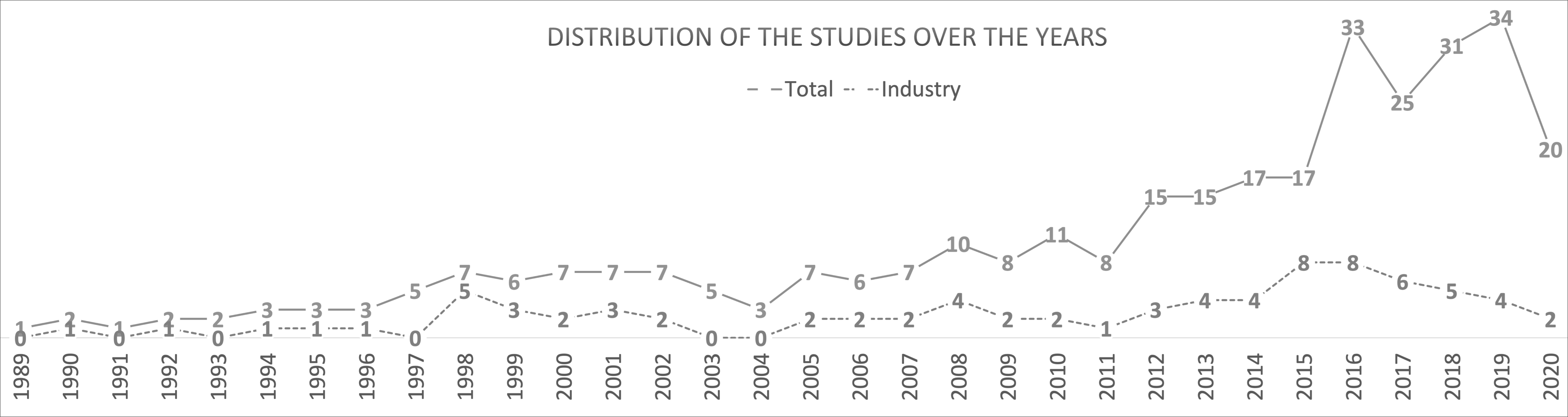}
\caption{Studies by year.}
\label{fig:year}
\end{figure}

\subsection{RQ1.1: Studies by Year}
\label{sec:year}

Fig.~\ref{fig:year} shows the number of publications per year, also considering industrial studies. We see that a slowly but steadily increasing number of works is available starting from 1989. Formal methods for railways therefore span over 30~years of research and applications. The number of papers increases from 2016 onwards, with a peak in 2016 and in 2019. As one can see from  Fig.~\ref{fig:year}, in 2016 publications almost double with respect to the previous two years, going from~17 to 33~papers. 

The overall increase since 2016 can be associated to the boost given to research in the railway domain by the Shift2Rail program (\url{https://shift2rail.org/}), which started providing grants exactly in 2016. The specific peak in 2016 is linked to the occurrence in the same year of the two conferences RSSRail (Int.\ Conf.\ on Reliability, Safety, and Security of Railway Systems: Modeling, Analysis, Verification, and Certification) and ISoLA (Int.\ Symp.\ on Leveraging Applications of Formal Methods, Verification and Validation). The former had its first edition in 2016, and it is specialized in railways, therefore being a natural venue for these types of studies. The second one is a bi-annual symposium, and regularly has a track dedicated to formal methods for railway systems. The peak in 2019 is again due RSSRail, and its co-occurrence with FMICS (Int.\ Conf.\ on Formal Methods for Industrial Critical Systems), a venue in which railway applications are a common topic. 

Concerning industrial studies, it is interesting to note that the first ones appear already in the early '90s. This indicates that formal methods for railways is born with a strong industrial focus, as industrially relevant problems have always been a primary concern. On the other hand, while for several years the proportion between industrial and academic studies is somewhat stable, we see that the radical increase in terms of papers observed in the last years is not followed by a corresponding abundance of industrial studies. On the contrary, industrial studies started decreasing after a peak in 2015--2016. This suggests that recent work mostly focuses on research problems, possibly exploiting new techniques, while industrial experimentation appears to be more limited\footnote{This could indicate that some formal approaches are now consolidated in industry. However, this conclusion cannot be drawn from the analysis, which focuses on research papers. A multi-vocal literature review would be needed to explore this.}.    

Information about the year of publication is used in the following to identify recent trends. Specifically, we consider a study to be \textit{recent} if it is published in the last 5~years, i.e., after~2015. This choice matches with the radical increase in~2016 in terms of number of publications in the field.

\begin{figure}[h]
\centering
\includegraphics[width=1\textwidth]{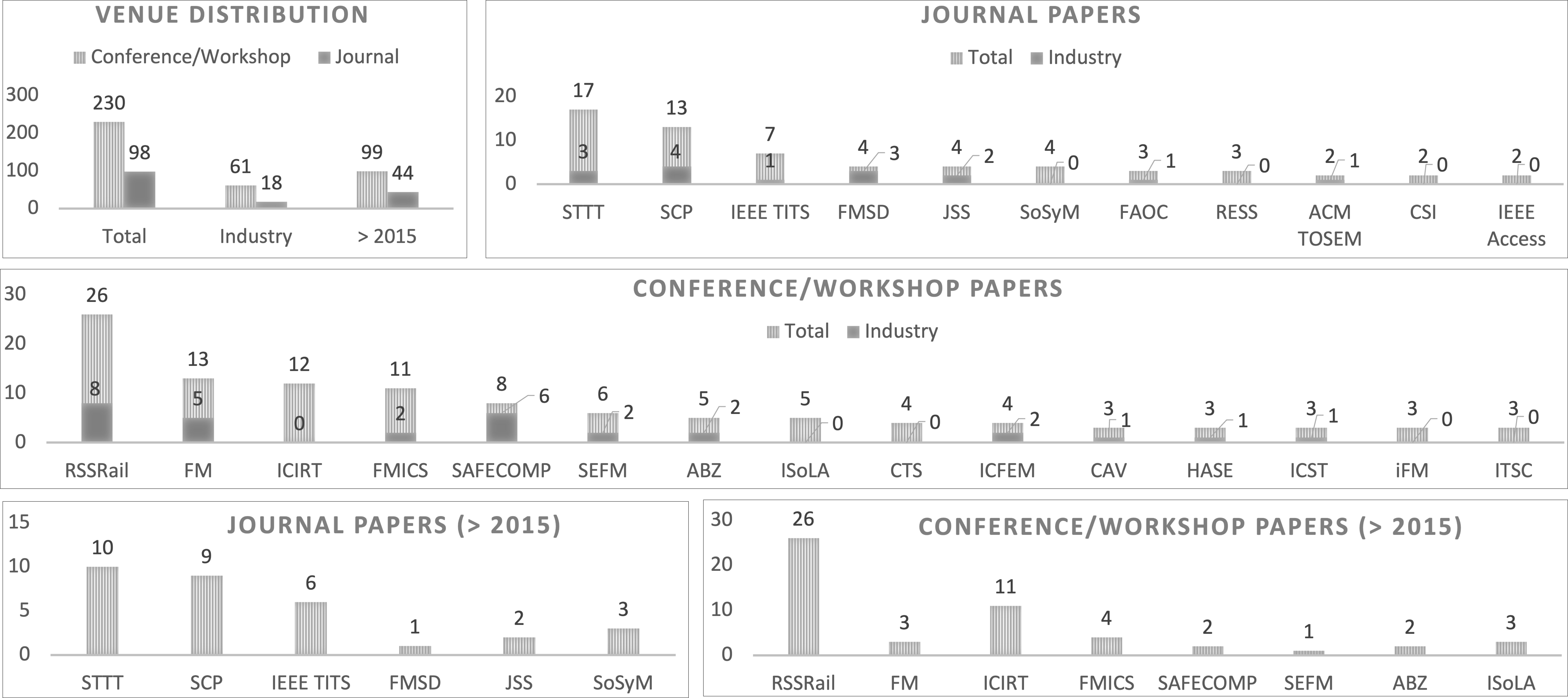}
\caption{Statistics on publication venues.}
\label{fig:venue}
\end{figure}

\begin{table}[]
\begin{footnotesize}
\begin{tabular}{|l|p{0.825\textwidth}|}
\hline
\multicolumn{2}{|c|}{\textbf{Journals}}\\ \hline
STTT & International Journal on Software Tools for Technology Transfer \\ \hline
SCP & Science of Computer Programming \\ \hline
IEEE TITS & IEEE Transactions on Intelligent Transportation Systems \\ \hline
FMSD & Formal Methods in System Design \\ \hline
JSS & Journal of Systems and Software \\ \hline
SoSyM & Software and Systems Modeling \\ \hline
FAOC & Formal Aspects of Computing \\ \hline
RESS & Reliability Engineering \& System Safety \\ \hline
ACM TOSEM & ACM Transactions on Software Engineering and Methodology \\ \hline
CSI & Computer Standards \& Interfaces \\ \hline
IEEE Access & IEEE Access \\ \hline
\hline
\multicolumn{2}{|c|}{\textbf{Conferences and Workshops}} \\ \hline
RSSRail & International Conference on Reliability, Safety and Security of Railway Systems \\ \hline
FM & International Symposium on Formal Methods \\ \hline
ICIRT & IEEE International Conference on Intelligent Rail Transportation \\ \hline
FMICS & international Conference on Formal Methods for Industrial Critical Systems \\ \hline
SAFECOMP & International Conference on Computer Safety, Reliability and Security \\ \hline
SEFM & International Conference on Software Engineering and Formal Methods \\ \hline
ABZ & International Conference on Rigorous State Based Methods \\ \hline
ISoLA & International Symposium on Leveraging Applications of Formal Methods, Verification and Validation \\ \hline
CTS & IFAC Symposium on Control in Transportation Systems \\ \hline
ICFEM & International Conference on Formal Engineering Methods \\ \hline
CAV & International Conference on Computer-Aided Verification \\ \hline
HASE & IEEE International Symposium on High Assurance Systems Engineering \\ \hline
ICST & IEEE International Conference on Software Testing, Verification and Validation \\ \hline
iFM & International Conference on integrated Formal Methods \\ \hline
ITSC & IEEE International Conference on Intelligent Transportation Systems \\ \hline
\end{tabular}
\end{footnotesize}
\caption{Description of acronyms for publication venues.}
\label{tab:acronyms}
\end{table}

\subsection{RQ1.2: Venue}
\label{sec:venue}

Fig.~\ref{fig:venue} reports the statistics on the venues in which papers about applications of formal methods to railways are published.  
The majority of the works are published in conferences~(230, 70\%), but a relevant percentage appears in journals~(98, 30\%). The distribution is the same for recent works. This indicates a well-established research field, with solid journal publications. On the other hand, the field is subject to ongoing development, with many conference and workshop contributions. The proportion leans towards conferences in a more marked way when considering industrial papers~(61, 77\% vs 18, 23\%). This can be linked to the tendency of companies to go for in-person dissemination venues, which can facilitate networking. Furthermore, journal publications may require disclosing more data, which is not always acceptable for company confidentiality policies.

Let us now look at the specific venues---for the sake of space, the plots report solely the most frequent ones. The acronyms of conferences/workshops and journals are described in Table~\ref{tab:acronyms}. 

Among \textbf{conference contributions}, RSSRail clearly dominates (26, 11\% of the conferences). This is not surprising, as this venue is specialized in rigorous methods applied to railway development. RSSRail is followed by FM~(13, 6\%), ICIRT~(12, 5\%) FMICS~(11, 5\%), SAFECOMP~(8, 3\%) and SEFM~(6, 3\%). FM is the flagship conference on formal methods, showing that the railway domain is particularly important for the whole community, and it is not a niche field of experimentation. The other venues are also not strictly focused on railways, but on intelligent transport systems, in the case of ICIRT, and on formal methods and software engineering applied to safety-critical systems. %, for the other venues. 
This landscape indicates that applications of formal methods to railway systems are considered relevant and well-accepted both in application-centered venues, like RSSRail and ICIRT, and in formal methods ones, like FM, FMICS, SAFECOMP and SEFM. Industrial works are particularly welcome in RSSRail, FM, FMICS and SAFECOMP. FM has an Industry Day forum organized in conjunction with the main symposium, which targets industrial development and use of formal methods. A selection of contributions to the Industry Day is published in the symposium proceedings.
Recent works confirm the historical landscape, although in this case specialized venues like RSSRail and ICIRT clearly outrank the others. This suggests that research efforts are now focused on applying existing formal methods, possibly tailoring them to specific railway applications. 

Considering \textbf{journal papers}, STTT~(17, 17\% of the journals) and SCP~(13, 13\%) are the most common venues, followed by IEEE TITS~(7, 7\%), FMSD~(4, 4\%), JSS~(4, 4\%) and~SoSym (4, 4\%). Special issues dedicated to FM and FMICS were published in STTT, SCP, FMSD and FAOC. Also industrial works are published in all these venues, except for IEEE TITS, and recent works follow the general trend. The considered journals are rather diverse in terms of focus. STTT is concerned with tools for technology transfer and the interplay between technology and industry. It is traditionally focused on formal tools and structured methods in general, and it is thus appropriate for papers that wish to experiment novel formal tools on railway systems. SCP, JSS and SoSym have a broader scope, more oriented to system modeling. IEEE TITS is specialized in transport systems, while FMSD is the only pure formal methods journal among those considered. %that is strictly concerned with formal methods. 
Overall, this landscape confirms the different interests of the research community towards the railway field, and that not only conferences but also a rather large spectrum of journals welcomes formal methods applied to railways. 

\begin{figure}[h]
\centering
\includegraphics[width=1\textwidth]{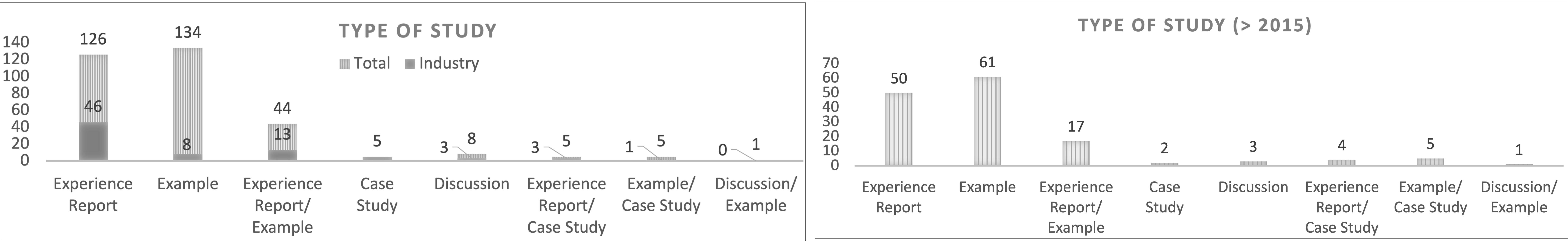}
\caption{Type of study.}
\label{fig:studytype}
\end{figure}

\subsection{RQ1.3: Empirical Evaluation}
\label{sec:studytype}

Fig.~\ref{fig:studytype} reports the statistics concerning the type of evaluation, considering the comparison between industrial studies and the total number of studies (top), and the recent trends (bottom).

The large majority of the studies are Examples~(134, 41\%), followed by Experience reports~(126, 38\%) and borderline cases between the two categories~(Experience Report/Example, 44, 13\%). The remaining papers concern Discussions~(8, 2.4\%), Case Studies~(5, 1.5\%) and other cases with less clear-cutting characterization. A different balance is identified for industrial studies, which are mostly Experience Reports~(46, 58\% of industrial studies) or other borderline cases in the same category~(Experience Report/Example, 13, 16\%; Experience Report/Case Study, 3, 4\%). All five Case Studies~(6\%) are industrial, as one expects from this type of research~\cite{runeson2012case}. These numbers indicate that most of the academic studies present Examples, to demonstrate or illustrate some formal technique. Instead, industrial studies tend to make a step forward and present real experiences. However, these experiences are mostly retrospective (i.e., Experience Reports) and do not concern the more mature form of Case Studies, with structured research questions and a rigorous process of data collection and analysis. 

The trends after~2015 (Fig.~\ref{fig:studytype}, bottom) match with the structure already observed for the whole set of studies. Hence, we argue that the focus on Examples and Experience Reports did not substantially change along the years. 

%\begin{figure}[h]
%\centering
%\includegraphics[width=1\textwidth]{./img-selected/venue.png}
% \caption{To-be-added: statistics about specific venues, and list of venues, 3 plots.}
% \label{fig:venue}
% \end{figure}

\begin{figure}[h]
\centering
\includegraphics[width=.6\textwidth]{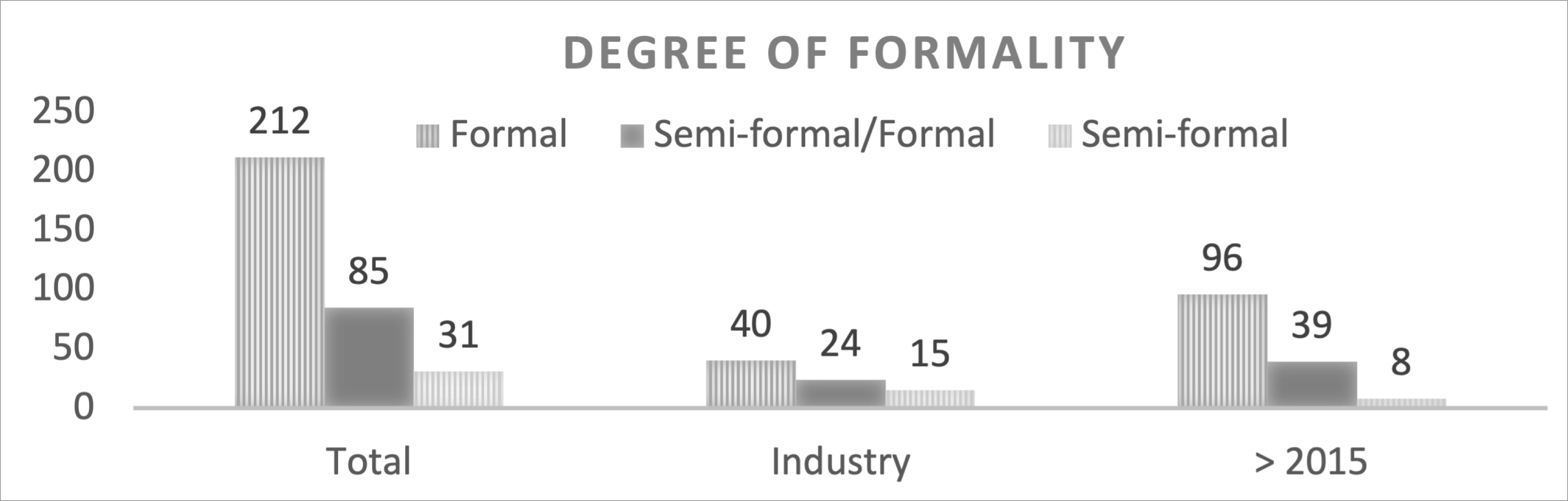}
\caption{Degree of formality.}
\label{fig:formality}
\end{figure}

\subsection{RQ2.1: Degree of Formality}
\label{sec:formality}

Fig.~\ref{fig:formality} reports the degree of formality of the techniques applied in the studies. Most of the works~(212, 65\%) are strictly Formal, part of them combine Formal and Semi-formal approaches~(85, 26\%) and the remaining ones are purely Semi-formal~(31, 9\%).

Interestingly, the proportion changes when considering solely industrial studies, for which half of the studies use exclusively Formal approaches~(40, 51\%), while the other half make use of Semi-formal techniques~(Semi-formal/Formal, 24, 30\%; Semi-formal, 15, 19\%). This suggests that industrial works tend to take into higher consideration Semi-formal approaches, arguably since these can help to bridge the gap between researchers and practitioners.

Recent works, instead, basically follow the general trend, with 96~Formal~(67\%), 39~Semi-formal/Formal~(27\%) and 8~Semi-formal~(6\%) studies. 

To summarize, Formal approaches dominate, with Semi-formal ones having a higher role in industrial studies. The trend did not substantially change in recent studies. 

\begin{figure}[h]
\centering
\includegraphics[width=1\textwidth]{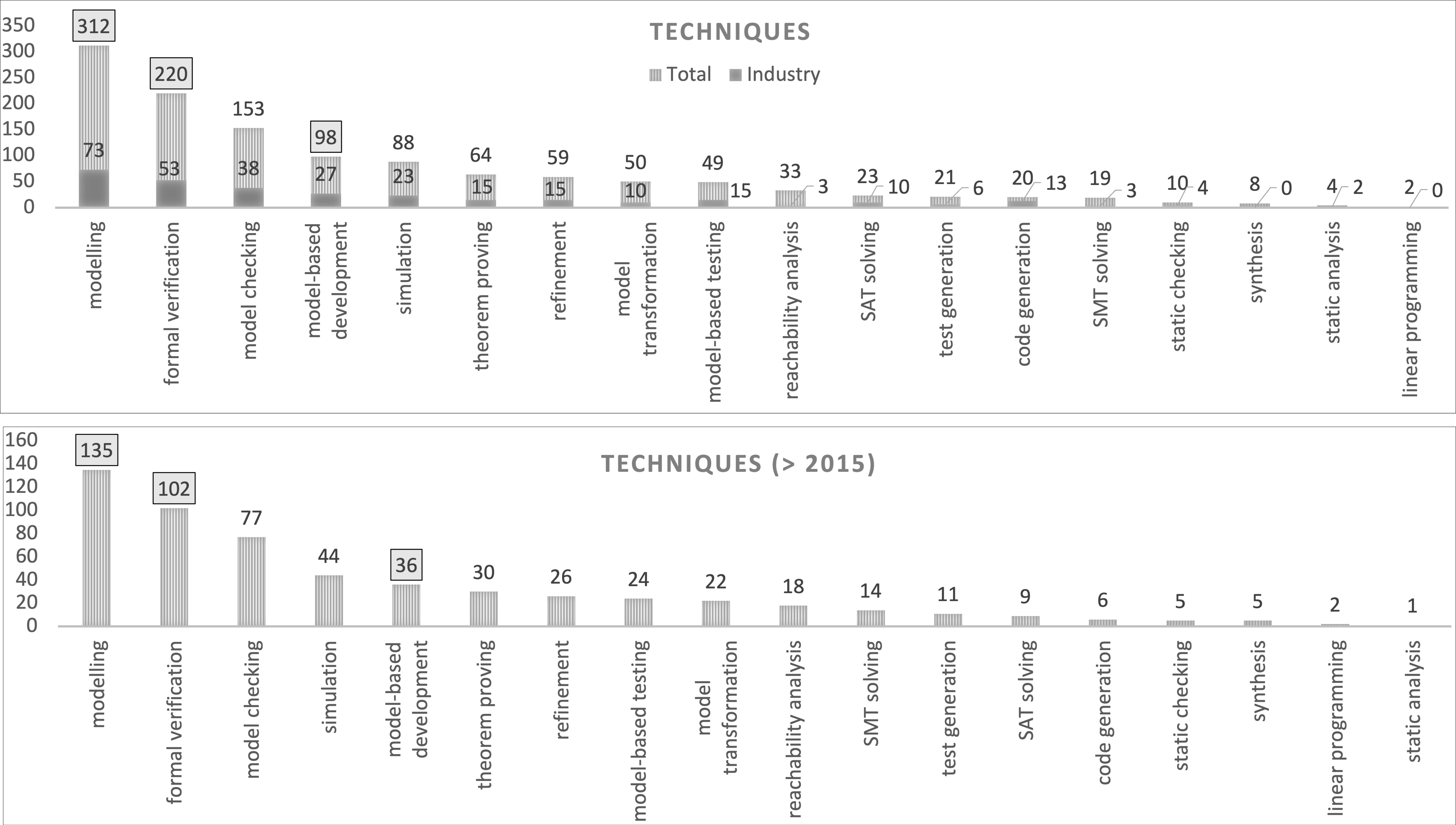}
\caption{Techniques.}
\label{fig:techniques}
\end{figure}

\subsection{RQ2.2: Techniques}
\label{sec:techniques}

Fig.~\ref{fig:techniques} reports the techniques used in the studies, according to the thematic analysis carried out. We report general families of techniques, namely modeling, formal verification and model-based development with highlighted labels.  
%It is worth noting that in this figure we do not consider hierarchical relations among the techniques, as these are entailed by the evaluation of the combination of techniques reported in Fig.~\cite{fig:combinationtechniques}, and discussed later. 

The vast majority of papers use the two fundamental techniques of formal methods, namely modeling~(312, 95\%) and formal verification~(220, 67\%). Though dominant, formal verification is not used in 33\%~of the studies, suggesting that other approaches, possibly non-formal, are used in combination with modeling.   
The most common technique for formal verification is model checking~(153, 47\%), used in about half of the works. The other classical verification techniques, namely theorem proving~(64, 19.5\%) and refinement~(59, 18\%) appear in a relevant, yet more limited number of studies. More frequent are other techniques such as the general family of model-based development~(98, 30\%) and simulation~(88, 27\%). The presence of other typical model-based techniques is also quite relevant, with model transformation~(50, 15\%), model-based testing~(49, 15\%) and reachability analysis~(33, 10\%) being frequently used. 

Techniques that are strictly related to code, like test generation~(21, 6\%), code generation~(20, 6\%) and static analysis~(4, 1\%) appear in a more limited number of papers. On the one hand, this variety of techniques indicates that railways is a playground for a large number of different approaches. On the other hand, this suggests that formal methods are typically applied on abstract, high-level models, and source code is only marginally considered.
Industrial studies seem to follow the same general trends, but with more attention to source code, as code generation and static analysis are used in over half of the studies~(13 out of~20 for code generation; and~2 out of~4 for static analysis). 

Recent studies, shown at the bottom of Fig.~\ref{fig:techniques}, indicate that the landscape is basically stable. However, some \rev{increasingly popular} techniques exist. In particular for simulation~(44, 31\% of recent studies), SMT solving~(14, 10\%),~model-based testing (24, 18\%) and test generation~(11, 8\%), half of the studies were published in the last five years. 

\begin{figure}[h]
\centering
\includegraphics[width=1\textwidth]{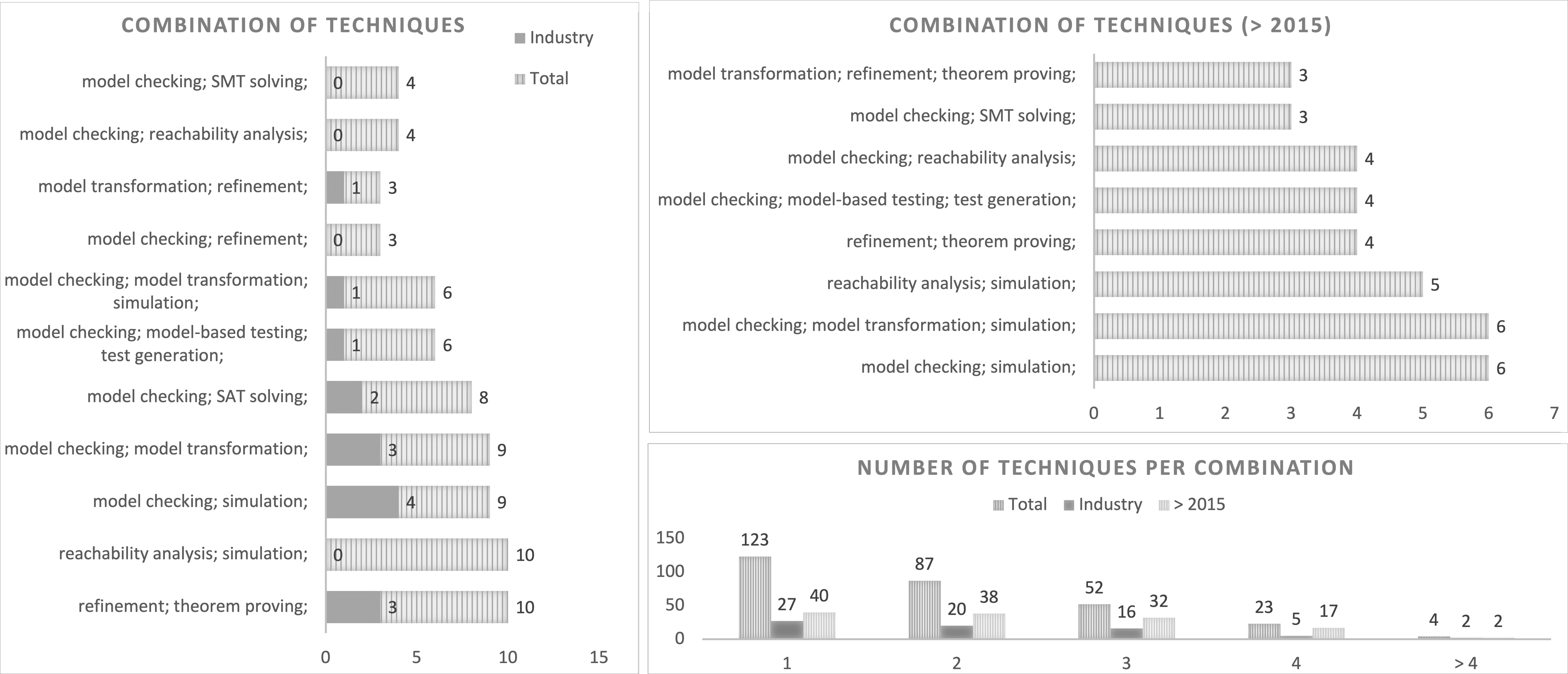}
\caption{Combination of techniques.}
\label{fig:combinationtechniques}
\end{figure}

Fig.~\ref{fig:combinationtechniques} shows the most frequent \textbf{combinations} of two or more techniques, and without considering the general families of modeling, formal verification and model-based development. Each combination is considered individually---subsets of combinations are not counted.

The bottom-right histogram indicates the number of techniques for each combination. We see that a large majority of the papers use only one technique~(123 papers in total, 38\%, 27 industrial, 34\%, 40 recent, 28\%)\footnote{The sum of papers does not amount to the total number of papers, as some of them used only modeling or only formal verification, without reference to known techniques.}, but a relevant number of papers use two to three techniques. The trend is similar in both industrial and recent papers, though recent papers appear to use also richer combinations (e.g., 12\% of the recent papers use 4~combinations, with respect to 7\%~of the total set). 

We now consider the specific combinations of techniques, by looking at the left and top-right histograms of Fig.~\ref{fig:combinationtechniques}. Here, we consider only combinations occurring in four or more papers (three or more for recent papers), to ease readability. Historically, the most frequent combination of techniques are \textit{refinement \& theorem proving}, \textit{reachability analysis \& simulation}, followed by model checking with other techniques. These other techniques include model transformation, simulation, SAT solving, model-based testing and test generation. In industrial papers, model checking occurs more frequently, in combination with other techniques in the model-based family. Interestingly, \textit{reachability analysis \& simulation}, a combination typically associated with Petri Nets, is never used in industrial papers, although it is the second one in terms of frequency. The greater relevance of model checking with respect to theorem proving is also visible in recent papers, in which \textit{model checking \& simulation}, and \textit{model checking \& model transformation \& simulation} are the most frequent combinations. Traditionally a combination of techniques matching the B~family (ProB in particular), we contribute this also to the recent popularity of applying statistical model checking (UPPAAL in particular) in the railway domain (cf.\ Sect.~\ref{sec:tools}). 

It is also worth noting that, out of 166 total papers that use two or more techniques, only 72 are represented in the plot (43\%). 
This indicates that a large majority of the papers use uncommon combinations and that a long tail of variants exists in these plots, as we will observe also for language families and tools in the next sections. 

\begin{figure}[h]
\centering
\includegraphics[width=1\textwidth]{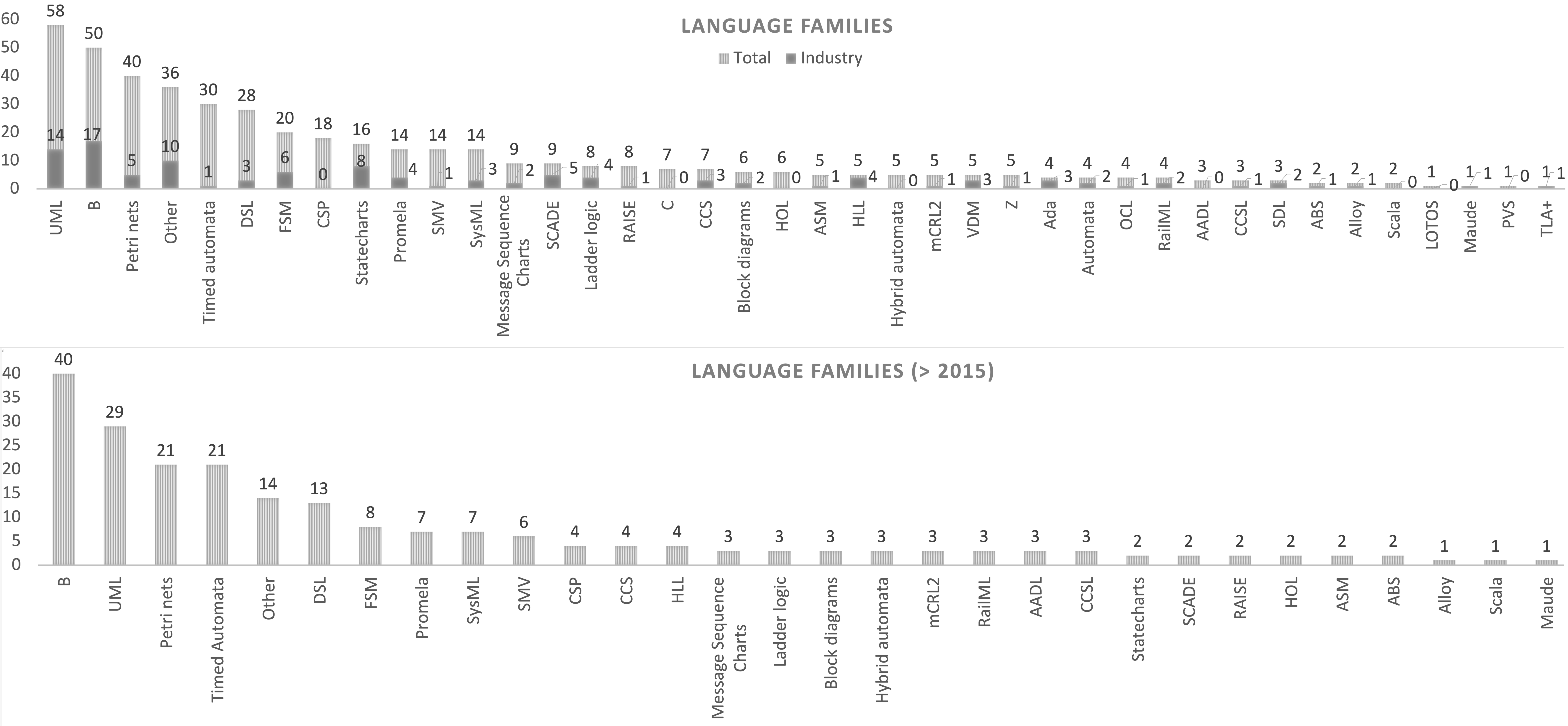}

\caption{Modeling language families considered in the studies.}
\label{fig:modelinglanguages}
\end{figure}

\subsection{RQ2.3: Language Families}
\label{sec:languagefamilies}
Fig.~\ref{fig:modelinglanguages} reports the statistics on the language families used in the studies. 
%This refers to the \textit{input} language used at the beginning of the process for modeling, and not to the languages that may be used later after some form of transformation. 
Dominant modeling language families are \rev{the semi-formal methods} UML~(58, 18\%) and Statecharts~(16, 5\%), \rev{the state-based formal methods} B~(50, 15\%), Timed automata~(30, 9\%) and Finite State Machines~(20, 6\%), \rev{the event-based formal methods} Petri nets~(40, 12\%) and CSP~(18, 5\%), and Domain-Specific Languages~(DSL, 28, 9\%). Other studies use tool-specific languages, such as Promela~(14, 4\%) and SMV~(14, 4\%). Besides these well-known language families, the plot shows a large number of languages that are used only in a limited number of studies---yet in many cases above~4. Furthermore, the placeholder \lq Other\rq, used for less established languages, appears as fourth most frequent language family, confirming that many works tend to be somehow unique, in terms of the language used. 

When looking at the number of industrial studies, the differences with respect to the general trend is rather evident. \rev{The state-based formal method} B~appears to be the most frequent modeling language family used in industrial works~(17, 22\%), followed by \rev{the semi-formal methods} UML~(14, 18\%) and Statecharts~(8, 10\%). Some languages appear to be used almost exclusively in academic works. These include Timed automata and \rev{the event-based formal methods} Petri nets and CSP. Others, instead, have a more industrial vocation, such as SCADE and High-level Language~(HLL), the input language of the SAT-based model checker S3~(Systerel Smart Solver).

Recent works also differ from the historical trends, with the B~language clearly occurring as the dominant one~(40, 20\%) and some modeling languages falling in the long tail, including industrially relevant languages like Statecharts~(2, 1\%). 

\begin{figure}[h]
\centering
\includegraphics[width=1\textwidth]{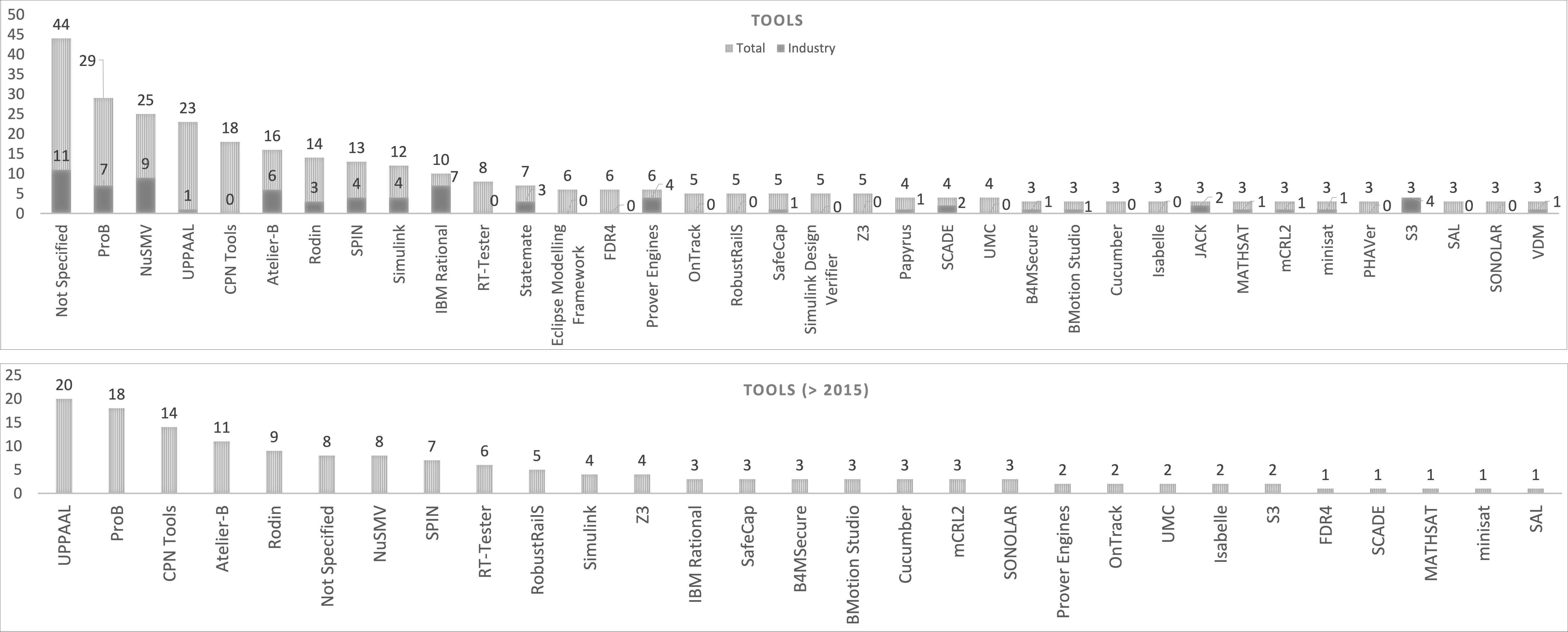}
\caption{Tools.}
\label{fig:tools}
\end{figure}

\subsection{RQ2.4: Tools}
\label{sec:tools}

Fig.~\ref{fig:tools} shows the results about the tools that are used in the papers. 
The majority of works do not indicate a specific tool~(Not Specified, 44, 13\% of the total). Frequently used tools are ProB~(29, 9\%), NuSMV~(25, 8\%), UPPAAL~(23, 7\%), CPN Tools~(18, 5\%), Atelier-B~(16, 5\%), Rodin~(14, 4\%), SPIN~(13, 4\%), Simulink~(12, 4\%), the IBM Rational family for UML and SysML~(10, 3\%), \textit{etc}. Tools in the B~family, namely ProB, Atelier-B and the Rodin platform clearly dominate, when considered together, but many other other well-known platforms are considered and a long tail of other tools, used solely in a few papers, can clearly be observed in the plot. 

Concerning industrial applications, we see that about the same percentage of papers does not specify a tool~(11, 14\%). The B~family still dominates, although NuSMV is the most frequently used tool in industrial works~(9, 11\%). Other tools in the long tail appear to have an industrial vocation, since the papers using them concern works with industry in more than half of the cases. These include Prover Engines, S3 and IBM Rational. Not surprisingly, these are closed-source tools that are not freely available, and experimentation in academia is naturally more oriented towards tools that have a free license and are extensible. %can be somehow extended. 
Other tools seem to be used almost exclusively in academic works in railways, namely CPN Tools~(0 industrial works out of 18) and UPPAAL~(1 out of 23). 

Recent works show a reduced tendency to have Not Specified tools. This suggests a greater attention in recent years to give importance to the tool used, and not only to the applied technique. The dominant toolset is UPPAAL~(20, 14\%), which includes UPPAAL SMC and UPPAAL Stratego, followed by ProB~(18, 13\%). Frequently used are also CPN Tools~(14, 10\%), Rodin~(11, 8\%), Atelier-B~(9, 6\%), NuSMV~(8, 6\%) and SPIN~(7, 5\%). This scenario indicates that UPPAAL is an \rev{increasingly popular tool}, although its usage in the railway industry is still limited, while ProB combines industrial uptake and frequency of use in recent works. The long tail of tools remains also for recent works, suggesting that the field is still a playground for experimentation with tools. Interestingly, many of these tools are specialized for railways, in particular RobustRailS, SafeCap and OnTrack. This suggests that while general purpose formal tools are used in the domain, there is a strong interest to tailor formal tools to the peculiarities of the domain. 

\begin{figure}[h]
\centering
\includegraphics[width=0.7\textwidth]{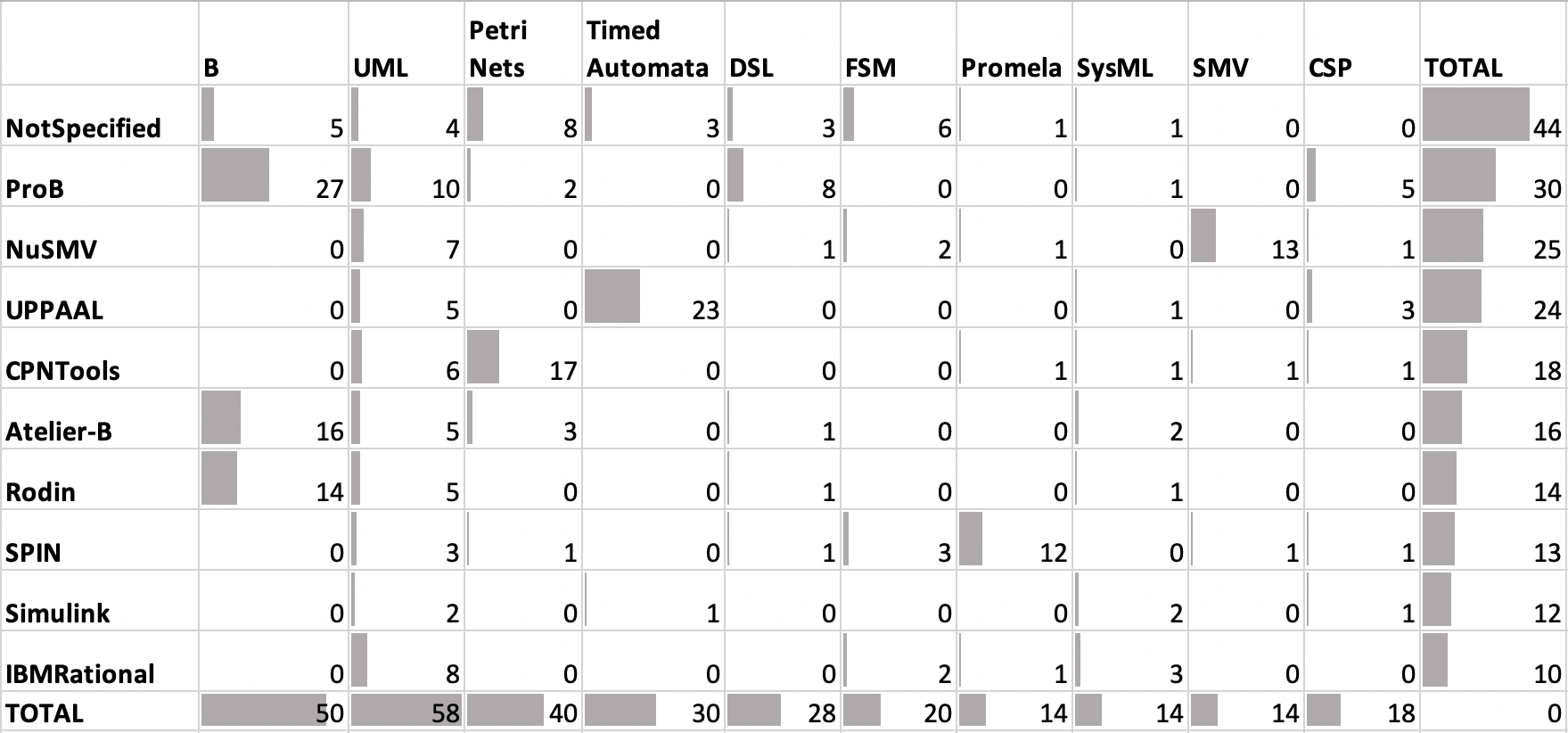}
\caption{Tools in relation to modeling languages}
\label{fig:toolsvslanguage}
\end{figure}

It is useful to look at the \textbf{relationships} between frequently used tools and modeling languages, reported in Fig.~\ref{fig:toolsvslanguage}. 
Besides the expected relationships between languages and tools, such as Promela for SPIN, SMV for NuSMV, Timed Automata for UPPAAL and B for ProB, Atelier-B and Rodin, there are some peculiar cases. In particular, the UML language is used in combination with all main tools, including ProB and NuSMV. Interestingly, with the exception of IBM Rational, none of the tools is specifically oriented to support UML. Thus, we conclude that UML is the language commonly used to model the system, but then the model is translated into the input language of different formal tools, e.g., to apply formal verification. This is in line with the fact that UML is the de facto industrial standard for documentation and communication among stakeholders. Another peculiar case is Petri nets, for which rather frequently the authors do not specify the support tool used. 

Finally, looking at Fig.~\ref{fig:toolsvslanguage}, it is also worth noting that some tools are used in papers in which different languages are used in combination. In particular, for many tools (e.g., ProB, SPIN, NuSMV) the sum of papers using them is smaller than the sum of the values that appear in their row cells. This phenomenon is less prominent for languages, where the total of papers is generally lower than the sum of the column cells. This suggests that a typical paper in our scope considers a single formal tool, but multiple modeling languages. While for ProB this is somehow in line with the vocation of the tool, which is oriented to be open to different input formalisms, for SPIN and NuSMV this can be related to the vocation of the tools as verification engines rather than design platforms, with limited graphical interfaces~\cite{ferrari2020comparing}, yet powerful formal verification capabilities.

\begin{figure}[h]
\centering
\includegraphics[width=.8\textwidth]{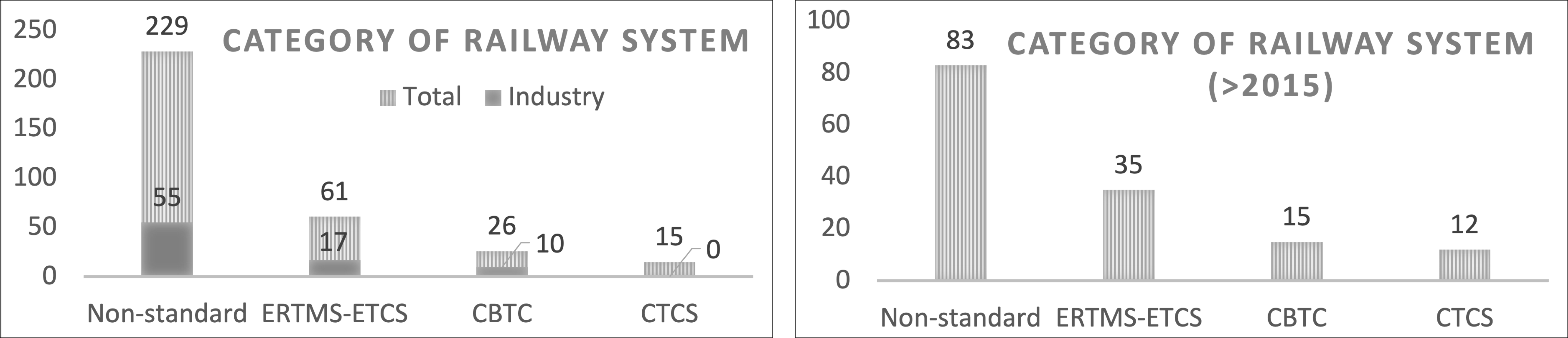}
\caption{Category of railway system.}
\label{fig:systemcategory}
\end{figure}

\subsection{RQ3.1: Category of Railway System}
\label{sec:systemcategory}

Fig.~\ref{fig:systemcategory} reports the distribution of system categories. 
The large majority of papers does not refer to any railway product standard~(229, 70\%), indicating that most
of the works focus on applications that either follow proprietary system specifications from some companies or 
are examples possibly inspired by real applications. A non-negligible number of works, however, is dedicated to the 
ERTMS-ETCS standard~(61, 19\%). This is followed by the CBTC~(26, 8\%) and the CTCS~(15, 5\%) standards. Industrial applications follow
the same trends, with slightly more applications using ERTMS-ETCS~(17, 22\% vs 19\%) and CBTC~(10, 13\% vs 8\%). 
Papers based on CTCS, instead, are not concerned with industrial applications. Considering recent studies, 
the percentage of works that focus on standards increases. In particular, although the majority of works is still classified as Non-standard~(83, 58\%), 
a slight increment is observed on works considering ERTMS-ETCS~(35, 24\% vs 19\%), CBTC~(115, 0\% vs 8\%) and CTCS~(12, 8\% vs 5\%). 

\begin{figure}[h]
\centering
\includegraphics[width=0.8\textwidth]{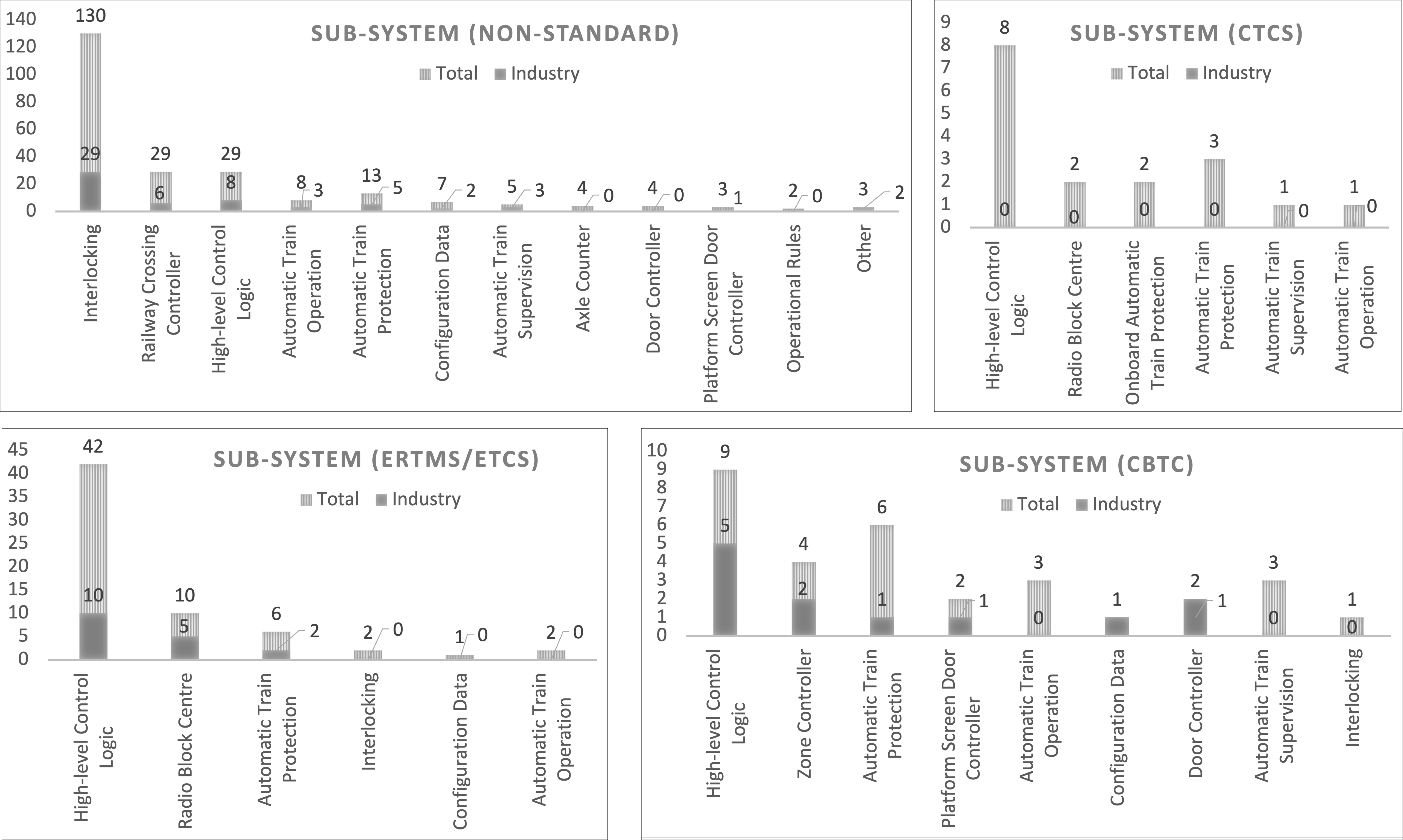}
\caption{Category of railway subsystem.}
\label{fig:subsystemcategory}
\end{figure}

\begin{figure}[h]
\centering
\includegraphics[width=1\textwidth]{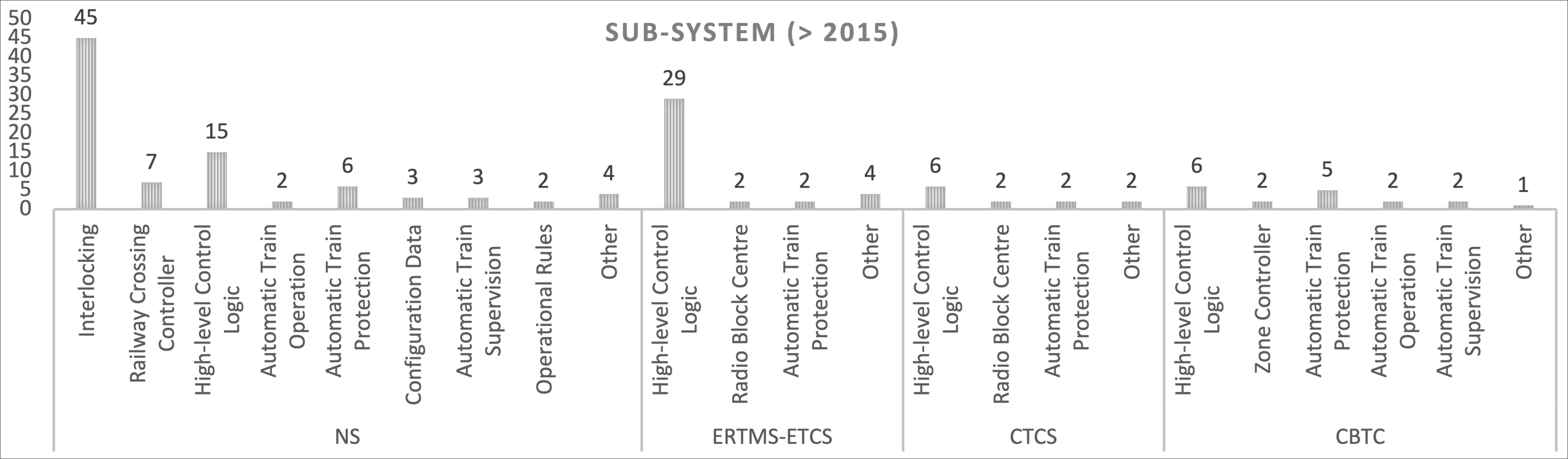}
\caption{Category of railway subsystem for recent papers.}
\label{fig:subsystemcategoryrec}
\end{figure}

\subsection{RQ3.2: Category and Railway Subsystem}
\label{sec:subsystemcategory}

Fig,~\ref{fig:subsystemcategory} reports the categories of subsystems, separated by category. The majority of non-standard subsystems are
Interlocking systems~(130, 40\%), followed by Railway Crossing Controllers~(29, 9\%) and High-level Control Logic~(29, 9\%). Then a large set of different subsystems
is covered, including ATO, ATP, %Automatic Train Operation, Automatic Train Protection, 
Configuration Data, etc. This indicates that formal methods have been applied
to a wide range of non-standard systems. The dominance of interlocking platforms is strictly linked to their equation-based tabular nature, which make them particularly amenable for formal verification by means of model checking or SMT solving. Interlocking platforms 
are also strongly present for industrial applications~(29, 37\%), confirming the prevalence of this type of subsystem in railway studies. 
High-level Control Logic, instead, is the typical (set of) subsystems considered in studies that use some standard: 42, 13\% for ERTMS; 9, 3\% for CBTC; 8, 2\% for CTCS. Overall, less variety in terms of subsystems types is observed for standardized cases, with the exception of CBTC, for which there is a higher balance in terms of different systems considered as one can visually grasp.

Looking at recent work in Fig.~\ref{fig:subsystemcategoryrec}, Interlocking still dominates, although in a less marked way~(45, 31\% vs 40\% for Non-standard systems). For the other subsystems the statistics are basically comparable with the historical ones, with the exception of Railway Crossing Controller~(7, 5\% vs 9\%). Apparently, the interest in this system, typically used in the past as an exemplary reference problem to experiment with new formal methods, tends to decrease in favor of other subsystems.

\begin{figure}[h]
\centering
\includegraphics[width=1\textwidth]{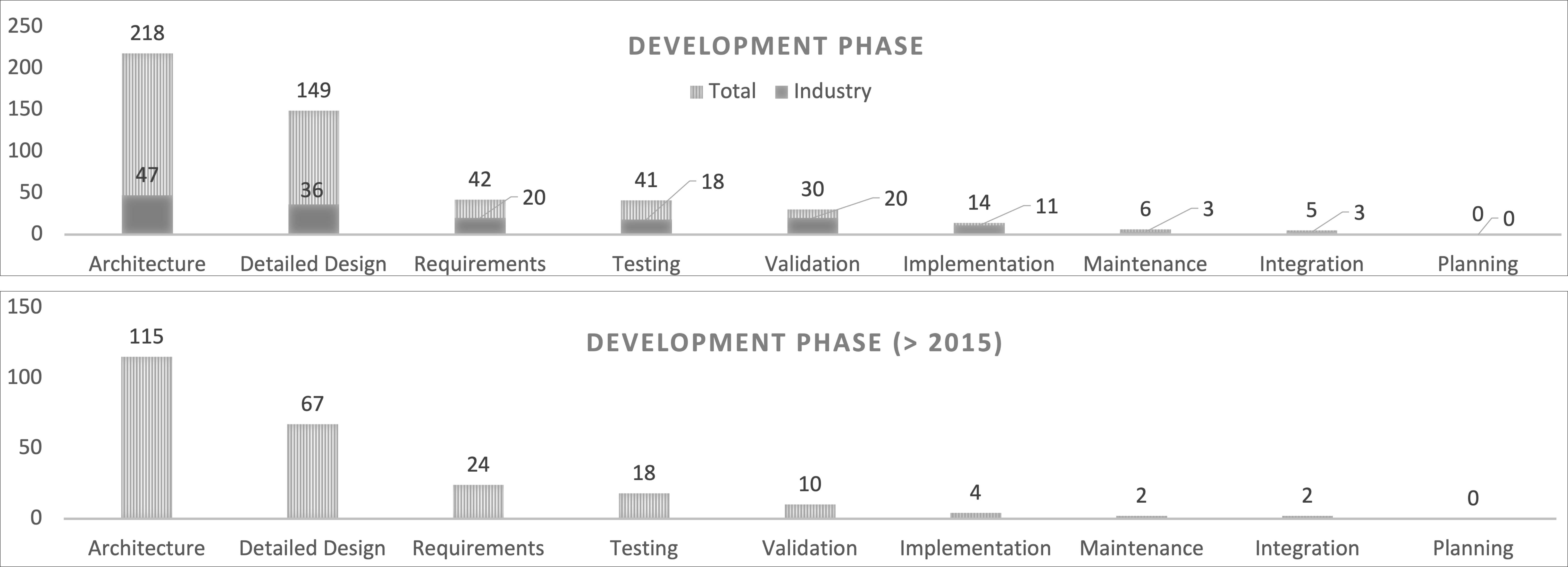}
\caption{Railway development phase.}
\label{fig:phase}
\end{figure}

\begin{figure}[h]
\centering
\includegraphics[width=0.525\textwidth]{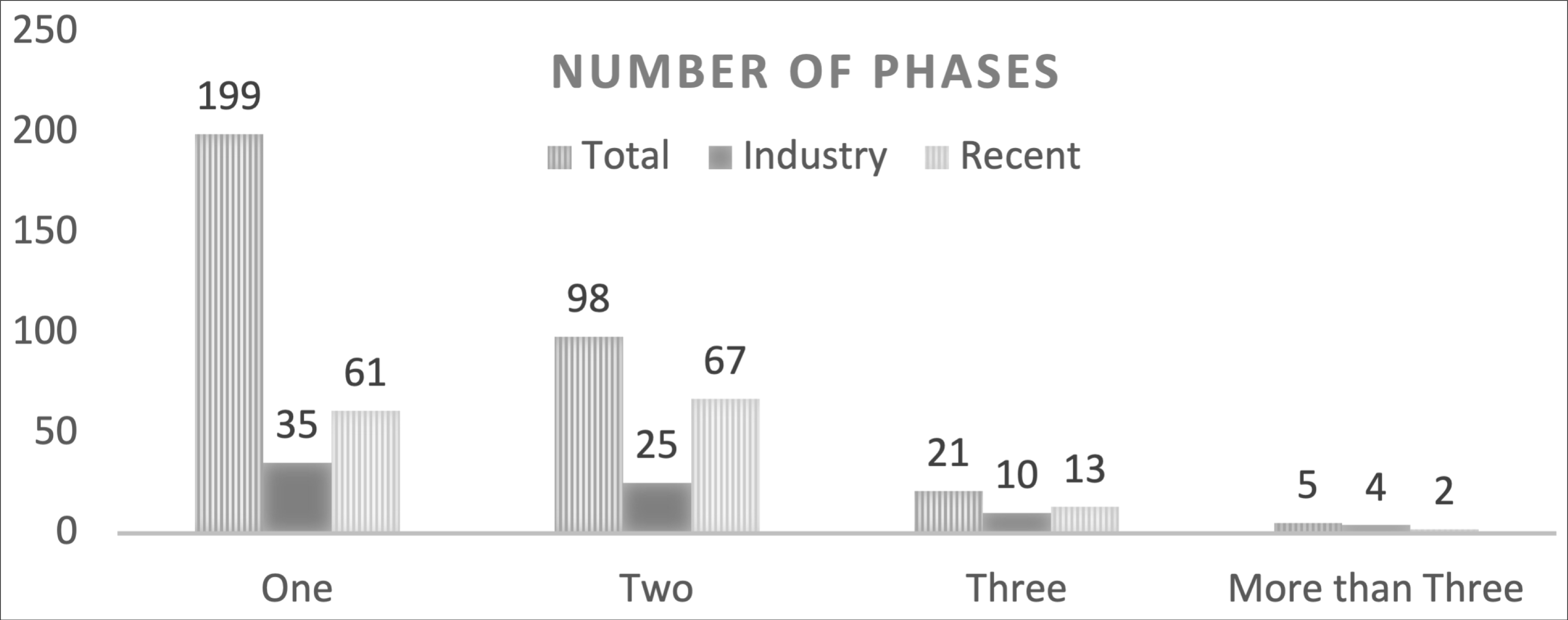}
\caption{Number of railway development phases.}
\label{fig:phasenumber}
\end{figure}

\begin{figure}[h]
\centering
\includegraphics[width=1\textwidth]{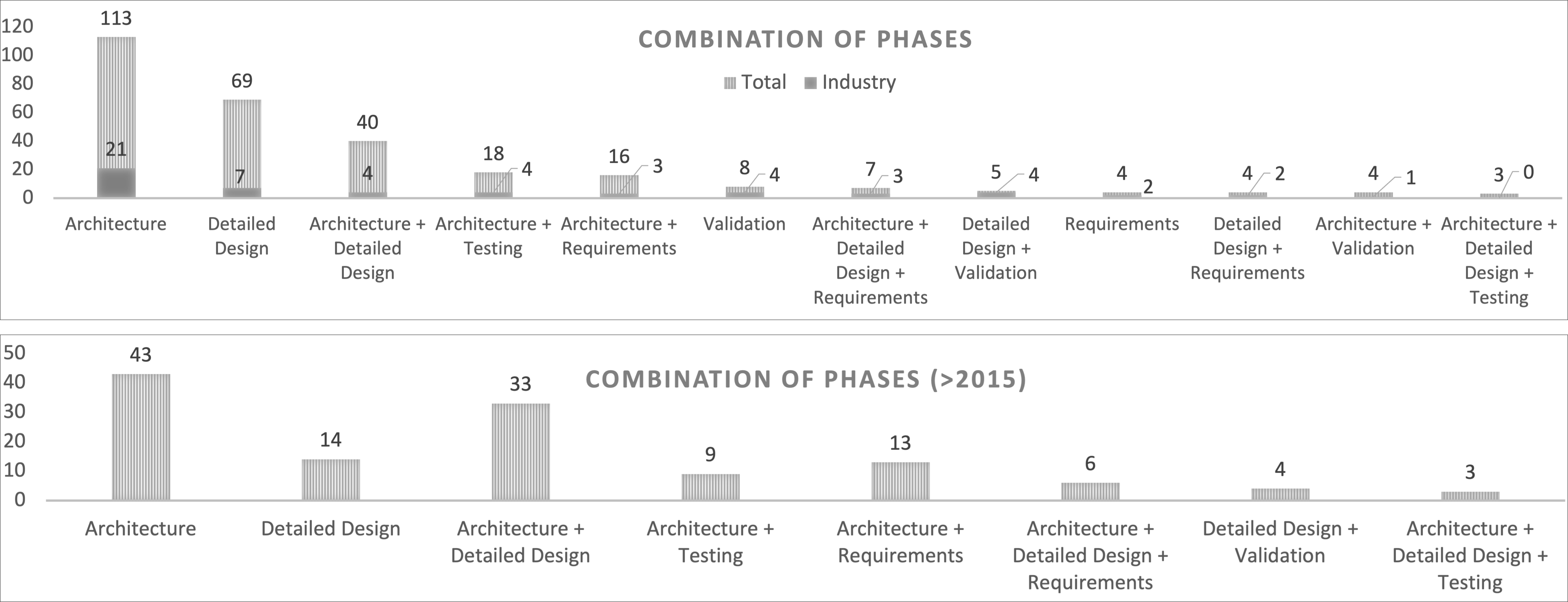}
\caption{Combination of railway development phases.}
\label{fig:phasecombo}
\end{figure}

\subsection{RQ3.3: Railway Development Phases}
\label{sec:phases}

Fig.~\ref{fig:phase} reports the development phases considered in the studies. The majority of them is concerned with Architecture~(218, 66\%) and Detailed Design~(149, 45\%), followed by Requirements~(42, 13\%), Testing~(41, 13\%) and Validation~(30, 9\%). This trend is followed by industrial studies, although these appear to be more focused on later phases of system development and on lower-level system representation. In particular, Validation and Implementation are considered in a relevant number of industrial studies (20 out of~30 for Validation, 11 out of~14 for Implementation). Furthermore, the percentage of studies focused on Architecture is lower for industrial studies with respect to the general trend~(47, 59\% vs 66\%). No notable differences can be observed for recent studies. 

Fig.~\ref{fig:phasenumber} shows the number of phases considered by each study. Most of the studies focus on one phase only~(199, 61\%), followed by two`(98, 30\%) and three~(21, 6\%) phases. A limited number of studies considers more than three phases. Industrial works tend to consider a higher number of phases, with 35~(44\% vs 61\%) focusing on one phase only, and 4~works covering more than three phases~(5\% vs 2\% of the total studies). The majority of recent works consider two phases~(67, 47\% vs 30\%) instead of one, marking a relevant difference with respect to the historical trends. 

Fig.~\ref{fig:phasecombo} reports the most frequent combinations of phases. Although many works focus solely on Architecture~(113, 34\%) and Detailed Design~(69, 21\%), several studies consider a combination of Architecture with other phases, namely Detailed Design~(40, 12\%), Testing~(18, 5\%) and Requirements~(16, 5\%). Industrial works that do not strictly focus on Architecture appear to be distributed over different combinations of phases, without a clear dominance. For recent works, Architecture + Detailed Design~(33, 23\%) directly follows Architecture alone~(43, 30\%) as typical combination. 

To summarize, the statistics show that formal methods have seen applications in almost all phases of railway system development with more focus on the design phases, namely Architecture and Detailed Design, also in combination. Industrial work tends to give more relevance to later phases, such as Implementation and Validation, and tends to consider a combination of a higher number of phases with respect to academic works.

\section{Summary and Discussion}
\label{sec:discussion}

In the following, we summarize the empirical findings of the study in relation to the main RQs, \rev{also pointing to representative papers}. For each question, we also discuss implications for research in the field of formal methods for railways.

%The most prominent empirical findings of our mapping study are the following. 
\smallskip

\textbf{RQ1. How is research demographically and empirically characterized in the field of applications of formal methods in the railway domain?} 

\smallskip

\begin{mdframed}
\textbf{Timeline.} Studies in formal methods for railways start in the late '80s, with a radical increase since 2016, thanks to the creation of dedicated venues (e.g., RSSRail) and the Shift2Rail program. 
\end{mdframed}

\smallskip

\begin{mdframed}
\textbf{Publication Venues.} 70\% of the works is published in conferences and 30\% in journals. Conferences are application-centered (RSSRail, ICIRT) as well as formal methods-centered (FM, FMICS, SAFECOMP, SEFM). Dominant journals are STTT and SCP. 
\end{mdframed}

\smallskip

\begin{mdframed}
\textbf{Evaluation.} The majority of studies are evaluated through Examples~(41\%) and Experience Reports~(38\%), while Case Studies are limited~(1.5\%)~\rev{\cite{LFFP11,ferrari13metro,BQJG15,HP16,CLMPM19}}.  
\end{mdframed}

\smallskip

\begin{mdframed}
\textbf{Industrial Involvement.} 68\%~of the studies have academic authors only, 8\%~have authors coming exclusively from industry and 24\%~have mixed affiliations. The majority~(68\%) considers industrial problems in laboratory settings, 16\%~validate the results with industrial partners and 5\%~document the development of real railway products with formal methods~\rev{\cite{BBFM99,BA05,LFFP11,ferrari2012lessons,ferrari13metro,HansenLKKNNSS20}}.
\end{mdframed}

\smallskip

Research in formal methods for railways has a solid tradition and several studies were published in collaboration with industrial partners. This indicates that formal techniques have a strong appeal for industries, and practitioners have interest in applying them to address problems that cannot be solved with other means. The presence of EU funding and dedicated venues clearly supports the development of research in the field. It is therefore advisable for researchers to take advantage of the current positive conjuncture, and 
%However, this should not imply that industrial relationships are neglected, as the decreasing trend in industrial works appears to show. 
make a step forward to better answer industrial demands by increasing the empirical rigor of their research. Despite the potential for sound industrial works, the empirical maturity of the field is still limited. Many works do not follow empirical standards, but simply report Examples or, in the best cases, retrospective Experience Reports. Based on this evidence, we argue that the community should attempt at answering existing questions with empirical software engineering lenses~\cite{wohlin2012experimentation}. This way, pressing questions, for which industry demands answers, can be addressed and the field can grow on the basis of scientific evidence. 
Research questions to address include the ones already discussed in previous work, also from other domains (e.g., aerospace and cybersecurity~\cite{mcdermid1998towards,michael2020open}), and revolve around the applicability of formal methods in real contexts, the maturity of tools~\cite{ferrari2020comparing,FMBB21}, their learning curve~\cite{steffen2017physics,garavel2019reflections}, their connection with the software engineering practice and processes~\cite{mcdermid1998towards,michael2020open,ferrari2012lessons} and how independent a company can realistically become from academic formal methods experts, e.g., through the usage of covert, hidden or lightweight formal methods~\cite{jackson2001lightweight}. These issues have been widely discussed in the literature, and appear to have not substantially changed over the years~\cite{GBP20}; \rev{an exception concerns cybersecurity: a large majority of experts recognises an important role for formal methods in cybersecurity. In the railway domain, however, cybersecurity is traditionally not considered as important as safety~\cite{VAAAM18} and the recently developed CENELEC technical specification~\cite{cenelec50701} for handling cybersecurity in the railway domain has yet to be transformed into a standard.}

Given the possibility offered by the strong industrial presence in the field, it is advisable to carry out research in the form of Case Studies, following established guidelines, like those by Runeson \textit{et al.}~\cite{runeson2012case}. Furthermore, not only Case Studies should be pursued, but also Laboratory Experiments, for example to compare software tools and evaluate user-related aspects. Quite surprisingly, our mapping study did not identify any form of controlled experiment. These are particularly common in software engineering~\cite{ko2015practical}, especially using students as subjects~\cite{daun2017experiments,falessi2018empirical}. A primary role here can be played by the community of formal methods education and training~\cite{dongol2019}. Specifically, by performing controlled experiments with students, instructors of formal methods can contribute not only to improving teaching practices, but also to the empirical assessment of formal methods. Overall, to advance the empirical maturity of the field, Experience Reports should become Case studies in the future, while Examples---which dominate the current literature, with novel formal approaches evaluated on limited cases---should become more sound Laboratory Experiments. 

We believe that carrying out more empirically sound studies may lead to publications outside the formal methods arena in leading software engineering venues, such as the Empirical Software Engineering journal, IEEE Transactions on Software Engineering and the ACM/IEEE Int.\ Conf.\ on Software Engineering (ICSE), where publications in formal methods and railways have already been published~\cite{chiappini2010,ferrari2020comparing}. This would give broader visibility to the formal methods community itself. 

\smallskip

\textbf{RQ2: What formal methods are used in the railway domain?}

\smallskip

\begin{mdframed}
\textbf{Formal vs Semi-formal.} Most of the studies are strictly formal~(65\%), while others use semi-formal methods~(9\%) or, more frequently, a combination of both~(26\%). 
\end{mdframed}

\smallskip 

\begin{mdframed}
\textbf{Techniques.} Formal modeling is applied in~95\% of the studies and formal verification in~67\%. Model checking is the most commonly adopted technique~(47\%)~\rev{\cite{DK01,chiappini2010,CRST11,LFFP11,CRST12,ferrari13metro,JamesMNRST14,BFBT16,BaoCZWMZ17,HHP17,BergerJLRS18,WuS18,CLMPM19,SS19,HansenLKKNNSS20,SnookHDFB21}}, followed by simulation~(27\%)~\rev{\cite{CCPRSTV99,LFFP11,ferrari13metro,FSZGLD14,BFBT16,BaoCZWMZ17,BergerJLRS18,wang2018hybrid,CLMPM19,SS19,HansenLKKNNSS20,SnookHDFB21}}, theorem proving~(19.5\%)~\rev{\cite{HL94,BBFM99,GM01,FurstHBSM16,CLMPM19}} and refinement~(18\%)~\rev{\cite{HL94,BA05,CRST11,LFFP11,JamesMNRST14,FurstHBSM16,HP16,HansenLKKNNSS20,SnookHDFB21}}. Less commonly used techniques are those strictly related to code, like test generation~(6\%)~\rev{\cite{ferrari13metro,BHHHPSH14,wang2018hybrid}}, code generation~(6\%)~\rev{\cite{BBFM99,BA05,LFFP11,ferrari2012lessons,ferrari13metro,SS19}} and static analysis~(1\%)~\rev{\cite{ferrari2012lessons,SS19}}. 38\%~of the papers use only one technique, while the rest uses combinations of two or more. Theorem proving in conjunction with refinement is the most frequent combination~\rev{\cite{HL94,FurstHBSM16}}.
\end{mdframed}

\smallskip

\begin{mdframed}
\textbf{Languages and Tools.} A large variety of modeling language families and tools is used. The dominant languages are UML~(18\%)~\rev{\cite{CRST11,CRST12,BQJG15,HP16,chiappini2010,WuS18,SnookHDFB21}} and B~(15\%)~\rev{\cite{BBFM99,BA05,LFFP11,AV13,JamesMNRST14,FurstHBSM16,CLMPM19,HansenLKKNNSS20,SnookHDFB21}}, while frequently used tools are ProB~(9\%)~\rev{\cite{LFFP11,AV13,JamesMNRST14,CLMPM19,HansenLKKNNSS20,SnookHDFB21}}, NuSMV~(8\%)~\rev{\cite{chiappini2010,CRST11,CRST12}} and~UPPAAL (7\%)~\rev{\cite{BaoCZWMZ17,wang2018hybrid}}. UML is normally used in combination with different formal tools. A typical paper considers a single formal tool, but multiple modeling languages. 
\end{mdframed}

\smallskip

The landscape of techniques, languages and tools is extensive. This confirms the findings of a previous questionnaire with railway stakeholders~\cite{BBFGMPTF18}, which highlighted the long tail of over 40~tools, even with only 44~respondents. On the one hand, this indicates that railways can be regarded as an appropriate field for research to experiment with a large variety of techniques, and confirms that this is a domain in which novel approaches can be tested. On the other hand, the fragmentation of techniques, languages and tools does not facilitate the work of practitioners, who face a paradox of choice when deciding what formal methods to adopt, as also observed by Steffen~\cite{steffen2017physics}. This is also not facilitated by the need to use combinations of techniques or tools, as done in part of the papers. There is therefore a need for a clearer classification of what techniques, languages and tools can and cannot do to facilitate the choice of practitioners. 

Despite this fragmentation, however, some latent patterns emerge, which deserve to be highlighted. UML is normally used for high-level representation, and models are normally translated into the input language of some formal verification engine. Typical choices are ProB, UPPAAL and NuSMV, which  cover quite diverse needs~\cite{ferrari2020comparing}, e.g., UPPAAL is appropriate when quantitative aspects come into play and when simulation is the best option; NuSMV can be chosen when complete state-space exploration is needed and the problem at hand can easily be represented as a state machine; ProB is recommended when prototyping, when an open platform is needed and also when one aims at top-down development of a monolithic system.   

Areas that need more exploration are also present, even in the wide landscape currently mapped. Specifically, research appears to neglect techniques that are closer to code, such as test generation, code generation and static analysis. 
Though it is widely believed since decades that formal methods and in particular formal verification techniques are at their best in the early design phases~\cite{Rus93,MWC10,GBP20}, it is the testing and debugging of the railway software that is the most resource consuming activity (about~50\% of the overall cost~\cite{myers2004art}) in safety-critical systems like railway systems. We thus encourage more research on applications of code-related formal methods, including software model checking and static analysis by means of abstract interpretation. 

\smallskip

\textbf{RQ3: In which way are formal methods applied to railway system development?}

\smallskip

\begin{mdframed}
\textbf{Systems.} 70\%~of the studies do not refer to any product standard, thus being either proprietary products or examples inspired by real cases. Product standards are considered in some cases, with ERTMS-ETCS~(19\%)~\rev{\cite{CCPRSTV99,chiappini2010,CRST11,CRST12,BHHHPSH14,HP16,HHP17,BergerJLRS18,HansenLKKNNSS20,SnookHDFB21}} and~CBTC (8\%)~\rev{\cite{LFFP11,wang2018hybrid,CLMPM19}} products. Most frequently considered systems are interlocking ones~\rev{\cite{BQJG15,BFBT16,HHP17,JamesMNRST14}}, and models of the high-level control logic describing the interaction of multiple subsystems~\rev{\cite{BBFM99,chiappini2010,CRST11,CRST12,FurstHBSM16,BergerJLRS18,SS19,HansenLKKNNSS20,SnookHDFB21}}. These are particularly common for standardized products.  
\end{mdframed}

\smallskip

\begin{mdframed}
\textbf{Phases.} The studies cover most of the railway development phases, with dominance of Architecture~(66\%)~\rev{\cite{HL94,BBFM99,CCPRSTV99,DK01,BA05,CRST11,ferrari2012lessons,AV13,ferrari13metro,BHHHPSH14,FSZGLD14,BQJG15,FurstHBSM16,HP16,BaoCZWMZ17,BergerJLRS18,wang2018hybrid,WuS18,SS19,HansenLKKNNSS20,SnookHDFB21}} and Detailed Design~(45\%)~\rev{\cite{BBFM99,BA05,ferrari2012lessons,ferrari13metro,JamesMNRST14,HHP17,CLMPM19,SS19,HansenLKKNNSS20,SnookHDFB21}}, followed by Requirements~(13\%)~\rev{\cite{BBFM99,GM01,chiappini2010,CRST12,ferrari13metro}}, Testing~(13\%)~\rev{\cite{ferrari2012lessons,ferrari13metro,BHHHPSH14,BFBT16,wang2018hybrid}} and Validation~(9\%)~\rev{\cite{BBFM99,LFFP11,ferrari2012lessons,ferrari13metro,BFBT16,CLMPM19,SS19}}. Most of the studies focus on only one phase~(61\%), followed by two (30\%) and three (6\%) phases. Architecture is frequently combined with other phases, namely Detailed Design~(12\%)~\rev{\cite{BBFM99,ferrari2012lessons,ferrari13metro,SS19,HansenLKKNNSS20,SnookHDFB21}}, Testing~(5\%)~\rev{\cite{BBFM99,ferrari2012lessons,AV13,ferrari13metro,BHHHPSH14,wang2018hybrid}} and Requirements~(5\%)~\rev{\cite{BBFM99,ferrari13metro}}.
\end{mdframed}

\smallskip

The limited consideration of product standards is correlated with the higher attention to interlocking products, which are typically not standardized. When standard systems are considered, works focus on the verification of their high-level control logic. This is in line with the needs of the railway infrastructure managers (e.g., RFI for Italy, SNCF for France), who need to ensure that the high-level specifications are satisfied by the products developed by different vendors~\cite{basile2020designing,BFR21}, so that they do not have to rely on a single provider. Nevertheless, formal methods are also needed for the providers themselves, as the CENELEC norms highly recommend their usage for the development of specific products~\cite{cenelec50128}. Furthermore, since the platforms that need to receive the certification are the individual subsystems (e.g., ATP, Axle Counters, etc.), more research should be dedicated to the application of formal methods to the verification and validation of single, standardized, subsystems.

The statistics also show that almost all the core railway development phases can be addressed with the support of formal methods, and this is in line with the recommendations of the norms~\cite{cenelec50128}. Nevertheless, additional effort should be dedicated to the later phases of the development process, and especially testing, implementation and validation, which are currently not sufficiently addressed. 

\smallskip

%ormal methods, and their usage for modeling and verification of high-level design and interfaces, can be particularly relevant to validate standard requirements specifications. Infrastructure managers  Therefore, the focus on  

\textbf{RQ-I: What are the characteristics of the studies reporting industrial applications?} 

\smallskip

\begin{mdframed}
\textbf{Demographics.} 
The total of 79~industrial studies represent~24\% of the whole corpus. Industrial studies are more frequently published in conferences \rev{than in journals}, and they are more frequently evaluated through Experience Reports~(58\% of industrial studies). 
\end{mdframed}

\smallskip

\begin{mdframed}
\textbf{Formal Methods and Techniques.}
Industrial studies follow the general trends for what concerns the usage of formal methods, with some differences. Specifically, the usage of semi-formal methods is more frequent in industrial studies \rev{with respect to academic ones}. In addition, studies that account for code-related aspects (i.e., using code generation or static analysis) often have some industrial involvement. The most frequent combination of techniques is \textit{model checking \& simulation}~(vs \textit{theorem proving \& refinement} for academic studies). 
\end{mdframed}

\smallskip

\begin{mdframed}
\textbf{Languages and Tools.}
B~is the most frequent modeling language family used~(22\%)~\rev{\cite{BA05,AV13,SnookHDFB21}}, followed by UML~(18\%)~\rev{\cite{chiappini2010,CRST12,BQJG15,HP16}} and Statecharts~(10\%)~\rev{\cite{CCPRSTV99,DK01}}. As for tools, those in the B~family dominate, although NuSMV is the most frequently used individual tool~(11\%)~\rev{\cite{chiappini2010,CRST12}}. Some closed-source tools have a clear industrial vocation (e.g., Prover Engines, S3 and IBM Rational). Others are applied almost exclusively in academic studies (e.g., CPN Tools and UPPAAL).
\end{mdframed}

\smallskip

\begin{mdframed}
\textbf{Railway Systems.} Industrial applications follow the same trends as academic ones, with slightly more applications using ERTMS-ETCS~(22\% vs 19\%)\rev{~\cite{chiappini2010,CRST12,HP16}} and CBTC~(13\% vs 8\%)\rev{~\cite{AV13}}. None of the industrial studies consider CTCS systems. Industrial works tend to give more relevance to later phases, such as Implementation and Validation, and tend to consider a combination of a higher number of phases with respect to academic works.
\end{mdframed}

\smallskip

Industrial works are a relevant part of the identified body of literature, which confirms the vocation of the field for industrial collaborations. Some aspects also indicate that industrial studies address issues that are less considered by academic ones, such as code-related techniques, later development phases and the reference to product standards. The main characteristics observed for the whole corpus also hold for industrial studies, and discrepancies are not substantial. One distinctive element, however, is the difference between tools with academic vocation and industrial ones. This implies that some tools, even widely used and industry-ready such as UPPAAL, are rarely used in railway-specific industrial works. We thus encourage researchers in formal methods to demonstrate the effectiveness of these tools in collaboration with railway partners. Furthermore, researchers should also consider experimenting with closed-source industrial tools like Prover Engines, S3 and the IBM Rational suite. While novel formal techniques can typically not be developed by researchers on top of these platforms, the evaluation of their usage in an industrial environment can highlight other process-related issues associated to the adoption of formal methods, and can open to further research opportunities for developers of academic tools.  

\smallskip

\textbf{RQ-T: What are the emerging trends of the last years?}

\smallskip

\begin{mdframed}
\textbf{Demographics.} 
There is a radical increase of studies post-2015, with a peak of 34~in~2019. The total of recent studies is~143 (44\% of all studies). After an increase also in industrial studies, recent years show a decline in favor of academic ones. Papers are mainly published in specialized application-oriented venues, like RSSRail and ICIRT. The historical trend of using Examples and Experience Reports as main evaluation methods did not change over the years. 
\end{mdframed}

\smallskip

\begin{mdframed}
\textbf{Formal Methods and Techniques.}
The landscape of techniques is stable, but some \rev{increasingly popular} techniques exist: for simulation~(31\% of recent studies)~\rev{\cite{BFBT16,BaoCZWMZ17,BergerJLRS18,wang2018hybrid,CLMPM19,SS19,HansenLKKNNSS20,SnookHDFB21}}, model-based testing~(18\%)~\rev{\cite{BFBT16,wang2018hybrid}}, SMT solving~(10\%)~\rev{\cite{HHP17}} and test generation~(8\%)~\rev{\cite{wang2018hybrid}}, half of the studies were published during the last 5~years. Recent works more frequently use complex combinations of techniques. 
\end{mdframed}

\smallskip

\begin{mdframed}
\textbf{Languages and Tools.}
In recent years, the B~language has taken the place of UML as the most common modeling language~(20\%)~\rev{\cite{FurstHBSM16,CLMPM19,HansenLKKNNSS20,SnookHDFB21}} and some languages fall in the long tail, including Statecharts~(1\% of recent works). \rev{Increasingly popular} tools are UPPAAL~(20, 14\%)~\rev{\cite{BaoCZWMZ17,wang2018hybrid}} and ProB~(13\%)~\rev{\cite{CLMPM19,HansenLKKNNSS20,SnookHDFB21}}. Many recently used tools are specialized for railways, e.g., RobustRailS, SafeCap and OnTrack.
\end{mdframed}

\smallskip

\begin{mdframed}
\textbf{Railway Systems.} The majority of the works is still classified as Non-standard~(58\%)~\rev{\cite{BQJG15,FurstHBSM16,WuS18,SS19}}, but a slight increment is observed on works considering the ERTMS-ETCS~\rev{\cite{BergerJLRS18,HansenLKKNNSS20}}, CBTC~\rev{\cite{CLMPM19}} and CTCS standards~\rev{\cite{BaoCZWMZ17}}. Interlocking is still the subject of the majority of the studies, but other subsystems (e.g., ATO, ATP and ATS)~\rev{\cite{HP16,wang2018hybrid,WuS18}} tend to be considered more frequently in recent years with respect to the past. Considered phases are in line with the historical trend, although recent works tend to address two phases instead of one only~(47\% vs 30\%). 
\end{mdframed}

\smallskip

The last 5~years see a rich amount of works, almost half of the total number of publications starting from~1989. These papers are characterized by a higher railway specialization, in terms of venues  and tools. This is in line with recommendations for the use of domain-specific formal methods already highlighted in the past~\cite{mcdermid1998towards}. Interestingly, recent works address some general shortcomings of preceding literature, like code-related aspects receiving more attention. The clear emergence of the use of tools like UPPAAL, together with the verification of non-safety critical railway systems like ATS and ATO, suggest a shift from the verification of the traditionally addressed safety problems to the verification of \textit{availability} problems, as previously recommended by Fantechi~\cite{Fan13}. 

What is worrying, however, is the decline of industrial studies in recent years. This may be due to the lower interest of industrial partners in the solutions offered by formal methods researchers, or to the stronger focus of academics on experimentation of more advanced techniques that are not industry-ready. In any case, we believe that the gap needs to be addressed to prevent disjuncture of the formal methods community from its traditional industrial connection.

% \begin{itemize}
%     \item Attention to problems that go beyond safety ones (non-safety related systems, as ATO and ATS, usage of UPPAAL)
%     \item Consideration of specialized tools for the domain, and specialized venues.
%     \item Consideration of tools that address the state-space explosion problem
%     \item these techniques are not yet applied in industrial works in railways.
%     inspiration shall be taken from other domains for UPPAAL.
%     \item interaction with industry is frequent, examples from industry are taken, and in some cases industrial works take the form of experience report, but there is no structured feedback loop. 
%     \item high-level models only, need for lower level ones.
%     \item higher attention to standards.
%     \item when FM are applied at the high level this is useful only for the infrastructure manager, does not help certification.
%     \item interplay between formal and semi-formal
%     \item variety of techniques, tools, languages
%     \item phases and railway certification
%     \item infrastructure managers and technology providers have different needs
%     \item open platforms as one size does not fit all and different tools perform different things. 
% \end{itemize}

\section{Threats to Validity}
\label{sec:threats}

We discuss the threats to validity of the current study and mitigation actions according to the categories identified by Ampatzoglou \textit{et al.}~\cite{AMPATZOGLOU2019201} for systematic reviews.

\paragraph{Study Selection Validity} The main threats to validity in this category are related to: (a)~the construction of the search string and its possibility to fail in identifying all relevant papers; (b)~the weaknesses of the search engines of the libraries used; (c)~the application of inclusion/exclusion criteria and quality criteria, which could be subjective. To address~(a), we piloted the string, and included a secondary search strategy through snowballing, which allowed us to identify additional papers not covered by the search string. To mitigate~(b), we performed the search multiple times, in three different rounds, and considering different engines. To mitigate~(c), we defined objective criteria based on previous work and piloted them, and when issues were identified, these were resolved through discussion among the authors. Furthermore, quality scores for each paper \rev{were} systematically cross-checked and discussed among the authors. \rev{A residual threat is the potential absence of relevant papers in the libraries, and the failure to identify them with snowballing. Given the set of mitigation measures, this threat should not substantially impact our results.} 

\paragraph{Data Validity} Major threats to data validity are: (a)~publication bias, as some applications of formal methods may have not appeared in research venues, \rev{also due to confidentiality issues of companies}; (b)~data extraction bias, due to possible subjectivity in data extraction; (c)~bias of the classification schemes, which are oriented to identify only specific data. Threats entailed by~(a) could not be mitigated entirely, although we argue that this issue is inherently reduced by the strong participation of industrial partners in the studies, and the presence of practitioner-oriented venues, such as RSSRail. To mitigate~(b), the data extractors, who have complementary competences, systematically cross-checked their results, and disagreement were addressed by involving a third expert subject. To address~(c), classification schemes were largely adapted from previous literature. \rev{Novel classification schemes introduced were defined after multiple iterations on samples of papers, so that they could be representative of the literature (cf. Sect.~\ref{sec:synthesis}). The schemes were piloted to ensure that they were correctly covering the content of the papers, using appropriate terminology. As the landscape is highly fragmented, we made an effort to keep an \textit{ad-hoc} degree of granularity, which could be representative of the papers that we reviewed. } 

\paragraph{Research Validity} To ensure research validity, we clearly reported the whole search and extraction process, and we shared the raw results our analysis, such that replication and independent analysis is possible. Research validity was further improved by the repetition of the process across three iterations, which confirmed that the adopted protocol can be replicated. 
\section{Conclusion}
\label{sec:conclusion}

This paper presents a systematic mapping study of applications of formal methods in the railway domain. We retrieve 328~high-quality studies published during the last 30~years, and we classify them considering their empirical maturity, the types of formal methods applied and railway specific aspects. Furthermore, we analyze recent trends and the characteristics of those studies that involve practitioners. Our results show that the field has a strong connection with the railway industry and research is currently thriving, with dedicated venues (RSSRail, ICIRT) and specialized tools (RobustRailS, SafeCap, OnTrack). We also identify a large and diverse set of languages, techniques and tools applied to different types of railway systems, highlighting the applicability of formal methods and tools and the suitability of the domain for the application of formal methods. 
On the other hand, we observe that the field needs to progress in terms of empirical maturity, as most of the published works are concerned with Examples or Experience Reports rather than more rigorous research efforts. Furthermore, we also notice that most of the research has so far focused on high levels of system abstraction and early development phases, while less work has been done in later railway phases, such as code and testing. Our work complements other empirical studies performed by the authors, which previously considered the perspective of stakeholders~\cite{BBFFGLM19,BBFGMPTF18} and surveyed different tools for railway system design~\cite{FMBB21,mazzanti2018towards,ferrari2020comparing}. This paper represents the cornerstone of this research endeavor oriented to present evidence concerning the state-of-the-art of formal methods in railways. As such, it provides a literature-based framework that can be used to understand and steer the research in the field, while facilitating further synergies with the railway industry. \rev{Large parts of our study protocol, and in particular the data extraction schemes, could also be adapted to other fields, like, e.g., automotive or avionics, so to provide a comparative analysis of the state of formal methods applications across different domains.}

\begin{acks}
This work has been partially funded by the ASTRail and 4SECURail projects.
These projects received funding from the Shift2Rail Joint Undertaking under the European Union's Horizon 2020 research and innovation programme under grant agreements No.\ 777561 (ASTRail) and 881775 (4SECURail).
The content of this paper reflects only the authors' view and the Shift2Rail Joint Undertaking is not responsible for any use that may be made of the included information.
\end{acks}

% Bibliography
\bibliographystyle{ACM-Reference-Format}
\bibliography{bibliography}

\end{document}